\documentclass{article}
\usepackage{amssymb}
\usepackage{amsfonts}
\usepackage{amsmath}
\usepackage{epsfig}
\usepackage{graphicx}

\begin{document}

\title{Space-time: Commutative or noncommutative ?}
\author{R. Vilela Mendes\thanks{%
rvilela.mendes@gmail.com; http://label2.ist.utl.pt/vilela/} \thanks{%
also at CMAFCIO, University of Lisboa} \\
%EndAName
IPFN, Instituto Superior T\'{e}cnico, University of Lisboa}
\date{ }
\maketitle

\begin{abstract}
Noncommutativity of the spacetime coordinates has been explored in several
contexts, mostly associated to phenomena at the Planck length scale.
However, approaching this question through deformation theory and the
principle of stability of physical theories, one concludes that the scales
of noncommutativity of the coordinates and noncommutativity of the
generators of translations are independent. This suggests that the scale of
the spacetime coordinates noncommutativity could be larger than the Planck
length. This paper attempts to explore the experimental perspectives to
settle this question, either on the lab or by measurements of phenomena of
cosmological origin.
\end{abstract}

\section{Introduction}

In this paper I will address the following questions:

"\textit{Is spacetime a commutative or a noncommutative manifold ?}"

"\textit{Can this question be decided in our time ?}", that is, are there
already sufficient experimental results and (or) experimental instruments to
decide ?

To sharpen these questions I will borrow from past results and from a few
new ideas. The emphasis will be on the experimental perspectives.

To my knowledge the first motivation to explore alternatives to the
continuous commutative spacetime manifold, was to cure the divergences
arising in the perturbative treatment of quantum fields. In this context
several discrete time and (or) discrete space models\ were proposed. However
these proposals violated Lorentz invariance and it was Snyder \cite{Snyder}
who made the first Lorentz invariant proposal%
\begin{equation}
\left[ x_{\mu },x_{\nu }\right] =\frac{ia^{2}}{\hbar }M_{\mu \nu }
\label{I1}
\end{equation}%
$M_{\mu \nu }$ being the Lorentz group generators. However the full Snyder
algebra lacked translation invariance and it was Yang \cite{Yang} who
pointed out that translation invariance would be recovered by interpreting
the coordinate operators as generators of linear transformations in
5-dimensional de Sitter space.

In recent years the noncommutativity of the spacetime coordinates, in the
sense%
\begin{equation}
\left[ x_{\mu },x_{\nu }\right] =i\theta _{\mu \nu }  \label{I2}
\end{equation}%
where $\theta _{\mu \nu }$ is either a c-number or an operator, has been
explored in many contexts (see for example \cite{Douglas}, \cite{Gaume}, 
\cite{Hinchliffe}, \cite{Hossenfelder} and references therein).
Noncommutative spacetime manifolds and noncommutative geometry techniques
appear naturally in the context of string and M-theory but, so far, they
lack a solid experimental or compelling physical motivation. \ An exception
might be the work in Refs. \cite{AhluwaliaPLB} \cite{Doplicher}. There it is
argued that attempts to localize events with extreme precision cause
gravitational collapse, so that spacetime below the Planck scale has no
operational meaning, leading to spacetime uncertainty relations. However,
this compelling reasoning would imply that the noncommutativity and the
associated fundamental length would be of the order of Planck's length $%
\lambda _{P}=\left( \frac{G\hbar }{c^{3}}\right) \simeq 1.6\times 10^{-33}$%
cm, far removed from current experimental reach. However, nothing forbids
that the noncommutativity of spacetime might occur at a bigger scale.
Nevertheless most recent discussions of noncommutativity of spacetime take
place in the framework of quantum gravity, therefore at the Planck length
scale (see for example the review \cite{Amelino} and references therein).

An alternative approach to the question of noncommutativity of spacetime is
based on deformation theory and the stability of physical theories.

\subsection{Noncommutative spacetime by deformation theory}

In the construction of models for the natural world, only those model
properties that are robust have a chance to be observed. It is unlikely that
properties that are too sensitive to small changes of the parameters will be
well described in the model. If a fine tuning of the parameters is needed to
reproduce some natural phenomenon, then the model is basically unsound and
its other predictions expected to be unreliable. For this reason a good
methodological point of view consists in focusing on the robust properties
of the models or, equivalently, to consider only models which are stable, in
the sense that they do not change, in a qualitative manner, when some
parameter changes. This is what has been called the \textit{stability of
physical theories principle }(SPTP) \cite{VilelaSPTP}.

The stable-model point of view led in the field of non-linear dynamics to
the rigorous notion of \textit{structural stability }\cite{Andronov} \cite%
{Smale}. As pointed out by Flato \cite{Flato} and Faddeev \cite{Faddeev} the
same pattern seems to occur in the fundamental theories of Nature. In
particular the passage from non-relativistic to relativistic and from
classical to quantum mechanics, may be interpreted as transitions from two
unstable theories to two stable ones. The stabilization of nonrelativistic
mechanics corresponds to the deformation of the unstable Galileo algebra to
the stable Lorentz algebra and quantum mechanics arises as the stabilization
of the Poisson algebra to the stable Moyal algebra. However, when the
generators of the Lorentz and the quantum mechanics Heisenberg algebra $%
\left\{ M^{\mu \nu },x^{\mu },p^{\mu }\right\} $ are joined together, one
finds out that the resulting Poincar\'{e}-Heisenberg algebra is also not
stable.

The Poincar\'{e}-Heisenberg algebra is deformed \cite{VilelaJPA} \cite%
{VilelaPLA1} to the stable algebra $\Re _{\ell ,\phi }=\{M^{\mu \nu },p^{\mu
},x^{\mu },\Im \}$ defined by the commutators%
\begin{equation}
\begin{array}{lll}
\lbrack M^{\mu \nu },M^{\rho \sigma }] & = & i(M^{\mu \sigma }\eta ^{\nu
\rho }+M^{\nu \rho }\eta ^{\mu \sigma }-M^{\nu \sigma }\eta ^{\mu \rho
}-M^{\mu \rho }\eta ^{\nu \sigma }) \\ 
\lbrack M^{\mu \nu },p^{\lambda }] & = & i(p^{\mu }\eta ^{\nu \lambda
}-p^{\nu }\eta ^{\mu \lambda }) \\ 
\lbrack M^{\mu \nu },x^{\lambda }] & = & i(x^{\mu }\eta ^{\nu \lambda
}-x^{\nu }\eta ^{\mu \lambda }) \\ 
\lbrack p^{\mu },x^{\nu }] & = & i\eta ^{\mu \nu }\Im \\ 
\lbrack x^{\mu },x^{\nu }] & = & -i\epsilon \ell ^{2}M^{\mu \nu } \\ 
\lbrack p^{\mu },p^{\nu }] & = & -i\epsilon ^{\prime }\phi ^{2}M^{\mu \nu }
\\ 
\lbrack x^{\mu },\Im ] & = & i\epsilon \ell ^{2}p^{\mu } \\ 
\lbrack p^{\mu },\Im ] & = & -i\epsilon ^{\prime }\phi ^{2}x^{\mu } \\ 
\lbrack M^{\mu \nu },\Im ] & = & 0%
\end{array}
\label{A1}
\end{equation}%
which, according to the SPTP paradigm, one would expect to be a more
accurate model. The stabilizing deformation introduces two new parameters $%
\ell ^{2},\phi ^{2}$ and two signs $\epsilon ,\epsilon ^{\prime }$. The
signs have physical relevance. For example, in the $\ell ^{2}\neq 0$ case if 
$\epsilon =+1$ time is discretely quantized and if $\epsilon =-1$ it will be
a space coordinate that has discrete spectrum.

An important point that this deformation \cite{VilelaJPA} of the Poincar\'{e}%
-Heisenberg algebra puts in evidence is the independence of the deformation
parameters $\ell $ (associated to the noncommutativity of the spacetime
coordinates) and $\phi $ (associated to the noncommutativity of momenta).

The stable algebra $\Re _{\ell ,\phi }=\left\{ x^{\mu },M^{\mu \nu },p^{\mu
},\Im \right\} $, to which the Poincar\'{e}-Heisenberg algebra has been
deformed, is isomorphic to the algebra of the $6-$dimensional
pseudo-orthogonal group with metric 
\begin{equation}
\eta _{aa}=(1,-1,-1,-1,\epsilon ,\epsilon ^{\prime }),\bigskip \ \epsilon
,\epsilon ^{\prime }=\pm 1  \label{A2}
\end{equation}%
Both $\ell $ and $\phi ^{-1}$ have dimensions of length. However they might
have different physical status and interpretation. Whereas $\ell $ might be
considered as a fundamental length and a new constant of Nature, $\phi ^{-1}$%
, being associated to the noncommutativity of the generators of translation
of the Poincar\'{e} group, is associated to the local curvature of the
spacetime manifold\footnote{%
In a de Sitter context $\phi $ would be the inverse of the (local) curvature
radius.} and therefore is a dynamical quantity related to the local
intensity of the gravitational field.

The two deformations, the one in the right-hand side of $[p^{\mu },p^{\nu }]$
and the one in the right-hand side of $[x^{\mu },x^{\nu }]$ are independent
from each other. Being associated to the local gravitational field, it is
natural that the scale of the deformation in the right-hand side of the $%
[p^{\mu },p^{\nu }]$ commutator be the Planck length scale ($10^{-33}$cm).
However there is no reason for the other one to have the same length scale.
A basic conjecture that will be explored in this paper is that $\ell $ is
much larger than $\phi ^{-1}$. In particular, a deformed tangent space would
correspond to take the limit $\phi ^{-1}\rightarrow \infty $ obtaining%
\begin{equation}
\left. \lbrack p^{\mu },p^{\nu }]\right\vert _{\phi ^{-1}\rightarrow \infty
}\rightarrow 0\hspace{2cm}\text{and\hspace{2cm}}\left. [x^{\mu },\Im
]\right\vert _{\phi ^{-1}\rightarrow \infty }\rightarrow 0  \label{A4}
\end{equation}%
all the other commutators being the same as in (\ref{A1}), leading to the
tangent space algebra $\Re _{\ell ,\infty }=\left\{ x^{\mu },M^{\mu \nu },%
\overline{p}^{\mu },\overline{\Im }\right\} $\footnote{$\overline{p}^{\mu },%
\overline{\Im }$ denote the tangent space ($\phi ^{-1}\rightarrow \infty $)
limits of the operators, not be confused with the physical $p^{\mu },\Im $
operators. According to the deformation-stability principle they are stable
physical operators only when $\phi ^{-1}$ is finite, that is, when gravity
is turned on.}.

The stable Poincar\'{e}-Heisenberg algebra in (\ref{A1}), obtained in \cite%
{VilelaJPA}, corresponds to a minimal deformation of the classical Poincar%
\'{e}-Heisenberg algebra. In \cite{VilelaJPA} it is also pointed out that
this deformation, not being unique, is the one that seems to be the most
reasonable one from a physical point of view. Chryssomalakos and Okon \cite%
{Chrysso} (see also \cite{Ahluwalia1} \cite{Ahluwalia2}) later careful
analysis has then found the structure of the most general deformations of
the Heisenberg-Poincar\'{e} algebra. This is summarized in the Appendix with
a critical analysis of the physical reasoning behind the choice of the
deformation in (\ref{A1}).

A first question of interest on the deformed algebras is the form of the
dispersion relations. For the deformed tangent space algebra $\Re _{\ell
,\infty }=\left\{ x^{\mu },M^{\mu \nu },\overline{p}^{\mu },\overline{\Im }%
\right\} $ it is%
\begin{equation*}
\left( \overline{p}^{0}\right) ^{2}-\left( \overline{p}^{1}\right)
^{2}-\left( \overline{p}^{2}\right) ^{2}-\left( \overline{p}^{3}\right) ^{2}=%
\overline{Q}_{2}
\end{equation*}%
the same as in the Poincar\'{e} algebra, because this algebra is unchanged
in $\Re _{\ell ,\infty }$, $\overline{Q}_{2}=m^{2}$ being the quadratic
Casimir operator.

For $\Re _{\ell ,\phi }=\left\{ x^{\mu },M^{\mu \nu },p^{\mu },\Im \right\} $
it is%
\begin{equation*}
P^{2}+\epsilon ^{\prime }\phi ^{2}\left( J^{2}-K^{2}\right) =Q_{2}
\end{equation*}%
with $P^{2}=p_{\mu }p^{\mu }$, $J^{i}=\frac{1}{2}\varepsilon ^{ijk}M_{jk}$, $%
K^{i}=M^{i0}$ and $Q_{2}$ is the quadratic Casimir operator for $SO\left(
3,2\right) $ ($\epsilon ^{\prime }=+1$) or $SO\left( 4,1\right) $ ($\epsilon
^{\prime }=-1$).

The fact that the right-hand-side of the commutator $[x^{\mu },x^{\nu }]$ is
a tensor operator rather than a c-number implies that most spacetime global
symmetries are preserved (see for example \cite{Morita}).

The deformed algebra (\ref{A1}) and its tangent space limit (\ref{A4}) have
far reaching consequences both for the geometry of spacetime \cite{VilelaJMP}
\cite{VilelaIJTP}, the dimension of the associated differentiable algebra,
the interactions of connection related quantum fields \cite{VilelaEPJ2} and
the Dirac equation \cite{VilelaMPLA}. Here however I will concentrate mostly
on possible experimental tests and estimates of the value of the deformation
parameters.

In the past, noncommutativity of the spacetime coordinates has been mostly
associated to quantum gravity effects and the Planck length scale. Although,
as pointed out in \cite{Amelino}, some particular physical situations might
greatly amplify the effects, the emphasis on the Planck length scale nature
of the noncommutativity has precluded the search for laboratory scale
effects. The point of view proposed in this paper is that the formal
independence of the deformation parameters $\ell $ and $\phi ^{-1}$ suggests
that these two length scales are naturally independent and therefore it
makes sense to look in the lab for the possibility of noncommutative effects
at a scale larger than the Planck length.

\section{Noncommutative spacetime: experimental perspectives}

From the commutator $[x^{\mu },x^{\nu }]=-i\epsilon \ell ^{2}M^{\mu \nu }$or
from a more general one , $\left[ x_{\mu },x_{\nu }\right] =i\theta _{\mu
\nu }$, one concludes that in the noncommutative case, the spacetime
coordinates cannot be treated in isolation and that at least an extra
operator is involved in all calculations in the spacetime manifold. In the $%
\epsilon =+1$ case the spacetime manifold is locally isomorphic to $SO(3,2)$
and in the $\epsilon =-1$ case to $SO(4,1)$. Convenient tools for
calculations are the representations of these algebras as operators on the
corresponding cones (see \cite{VilelaJMP} and the appendixes in \cite%
{VilelaEPJ1} and \cite{VilelaIJTP}), irreducible representations of these
algebras playing the role of "points" in their noncommutative geometry.

Here one analyses a few situations were the noncommutativity of spacetime
might be tested and measured as well as some of the instances where such
tests seem at present to be unfeasible. When the nature of the
noncommutativity is left essentially unspecified, as in $\left[ x_{\mu
},x_{\nu }\right] =i\theta _{\mu \nu }$, it is difficult to obtain clearly
testable predictions. Therefore here, as a working principle, use will
always be made of the commutation relations in (\ref{A1}), in particular in
the tangent space limit (\ref{A4}).

\subsection{Measuring speed}

In the noncommutative context, space and time being noncommutative
coordinates, they cannot be simultaneously diagonalized and speed can only
be defined in terms of expectation values, that is%
\begin{equation}
v_{\psi }^{i}=\frac{1}{\left\langle \psi _{t},\psi _{t}\right\rangle }\frac{d%
}{dt}\left\langle \psi _{t},x^{i}\psi _{t}\right\rangle  \label{2.1}
\end{equation}%
where $\psi $ is a state with a small dispersion of momentum around a
central value $p$. At time zero%
\begin{equation}
\psi _{0}=\int \left\vert k^{0}\overset{\longrightarrow }{k}\alpha
\right\rangle f_{p}\left( k\right) d^{3}k  \label{2.2}
\end{equation}%
with $k^{0}=\sqrt{\left\vert \overset{\longrightarrow }{k}\right\vert
^{2}+m^{2}}$, $\alpha $ standing for the quantum numbers associated to the
little group of $k$ and $f_{p}\left( k\right) $ is a normalized function
peaked at $k=p$.

In \cite{VilelaPLA3} a first order (in $\ell ^{2}$) derivation of the speed
corrections was obtained. Here a more complete treatment will be done. To
obtain $\psi _{t}$ one applies to $\psi _{0}$ the time-shift operator, which
is not $e^{-iap^{0}}$\ because%
\begin{equation}
e^{-iap^{0}}te^{iap^{0}}=t+a\Im  \label{2.3}
\end{equation}%
follows from 
\begin{equation}
\left[ p^{0},t\right] =i\Im  \label{2.4}
\end{equation}%
whereas a time-shift generator $\Upsilon $ should satisfy%
\begin{equation}
\left[ \Upsilon ,t\right] =i\mathbf{1}  \label{2.5}
\end{equation}%
Here the calculations are carried out in the $\Re _{\ell ,\infty }$ algebra.
To implement the commutation relations of the deformed tangent space algebra 
$\Re _{\ell ,\infty }=\left\{ x^{\mu },M^{\mu \nu },\overline{p}^{\mu },%
\overline{\Im }\right\} $, use a basis where the $5$-variables set $\left( 
\overline{p}^{\mu },\overline{\Im }\right) $ is diagonalized\footnote{%
Notice that it is only in the tangent space algebra $\Re _{\ell ,\infty }$
that the operators $\left( \overline{p}^{\mu },\overline{\Im }\right) $ may
be simultaneously diagonalized, not in the full algebra $\Re _{\ell ,\phi }$.%
}. In this basis the commutation relations are realized by%
\begin{eqnarray}
x^{\mu } &=&i\left( \epsilon \ell ^{2}\overline{p}^{\mu }\frac{\partial }{%
\partial \overline{\Im }}-\overline{\Im }\frac{\partial }{\partial \overline{%
p}_{\mu }}\right)  \notag \\
M_{\mu \nu } &=&i\left( \overline{p}_{\mu }\frac{\partial }{\partial 
\overline{p}^{\nu }}-\overline{p}_{\nu }\frac{\partial }{\partial \overline{p%
}^{\mu }}\right)  \label{2.6}
\end{eqnarray}

\bigskip Then, one obtains the following time shift operator $\Upsilon $ in (%
\ref{2.5}), to all $\ell ^{2}$ orders%
\begin{equation}
\Upsilon =\frac{\overline{p}^{0}}{\overline{\Im }}\sum_{k=0}\left( -\epsilon
\right) ^{k}\frac{\ell ^{2k}}{2k+1}\left( \frac{\overline{p}^{0}}{\overline{%
\Im }}\right) ^{2k}  \label{2.7}
\end{equation}%
To obtain this result, use may be made of $\left[ t,\overline{\Im }^{-1}%
\right] =-i\epsilon \ell ^{2}\overline{p}^{0}\overline{\Im }^{-2}$, which
follows from $\left[ t,\overline{\Im }\overline{\Im }^{-1}\right] =0$ and $%
[t,\overline{\Im }]=i\epsilon \ell ^{2}\overline{p}^{0}$. Alternatively one
may check that (\ref{2.7}) satisfies (\ref{2.5}) using the representation (%
\ref{2.6}) to obtain%
\begin{equation*}
\left[ \Upsilon ,x^{0}\right] =i\sum_{k=0}\left\{ \left( -\epsilon \right)
^{k}\ell ^{2k}\left( \frac{\overline{p}^{0}}{\overline{\Im }}\right)
^{2k}-\left( -\epsilon \right) ^{k+1}\ell ^{2k+2}\left( \frac{\overline{p}%
^{0}}{\overline{\Im }}\right) ^{2k+2}\right\}
\end{equation*}

More compact forms of the time-shift operator are%
\begin{equation}
\Upsilon =\left\{ 
\begin{array}{ccc}
\frac{1}{\ell }\tan ^{-1}\left( \ell \frac{\overline{p}^{0}}{\overline{\Im }}%
\right) &  & \epsilon =+1 \\ 
\frac{1}{\ell }\tanh ^{-1}\left( \ell \frac{\overline{p}^{0}}{\overline{\Im }%
}\right) &  & \epsilon =-1%
\end{array}%
\right.  \label{2.7a}
\end{equation}%
Now one computes the time derivative of the expectation value of $x^{i}$ on
the time-shifted state%
\begin{equation}
\psi _{t}=\int \exp \left( -it\Upsilon \right) \left\vert \widetilde{k}^{0}%
\widetilde{k}^{i}\alpha \right\rangle f_{p}\left( \widetilde{k}\right) d^{3}%
\widetilde{k}  \label{2.8}
\end{equation}%
From (\ref{2.6}) and (\ref{2.7a}) one has%
\begin{equation*}
x^{i}e^{-it\Upsilon }=e^{-it\Upsilon }t\frac{\overline{p}^{i}}{\overline{p}%
^{0}}\frac{1-\epsilon \ell ^{2}\left( \frac{\overline{p}^{0}}{\overline{\Im }%
}\right) ^{2}}{1+\epsilon \ell ^{2}\left( \frac{\overline{p}^{0}}{\overline{%
\Im }}\right) ^{2}}
\end{equation*}%
Therefore the wave packet velocity is

\begin{equation}
v_{\psi }=\frac{\overline{p}}{\overline{p}^{0}}\frac{1-\epsilon \ell
^{2}\left( \frac{\overline{p}^{0}}{\overline{\Im }}\right) ^{2}}{1+\epsilon
\ell ^{2}\left( \frac{\overline{p}^{0}}{\overline{\Im }}\right) ^{2}}
\label{2.9}
\end{equation}%
a result that holds to all $\ell ^{2}$ orders in $\Re _{\ell ,\infty }$. In
leading order it is $v_{\psi }\simeq \frac{\overline{p}}{\overline{p}^{0}}%
\left( 1-2\epsilon \ell ^{2}\left( \frac{\overline{p}^{0}}{\overline{\Im }}%
\right) ^{2}\right) $. Notice that the correction is negative or positive
depending on the sign of $\epsilon $. For example, a massless particles wave
packet would be found to travel slower or faster than $c$ according to
whether $\epsilon =+1$ (quantized time) or $\epsilon =-1$ (quantized space).
Also notice that this deviation from $c$, for the velocity of the massless
particle wave packet, implies no violation of relativity. Both the Lorentz
and the Poincar\'{e} groups are still exact symmetries in $\Re _{\ell
,\infty }$ and the velocity corrections do not arise from modifications of
the dispersion relation for elementary states, which still is%
\begin{equation}
\left( \overline{p}^{0}\right) ^{2}=\left( \overrightarrow{\overline{p}}%
\right) ^{2}+m^{2},  \label{2.10}
\end{equation}%
but from the noncommutativity of time and space.

Now some of the existing experimental results will be analyzed to find
bounds on the value of $\ell $ (a fundamental time or a fundamental length).

In the corrected 2012 OPERA\ data \cite{OPERA} for 17 GeV neutrinos, the
reported result is%
\begin{equation}
\left\vert \frac{v-c}{c}\right\vert =\left( 2.7\pm 3.1\left( \text{stat}%
\right) 
\begin{array}{c}
+3.4 \\ 
-3.3%
\end{array}%
\left( \text{sys}\right) \right) \times 10^{-6}  \label{2.11}
\end{equation}%
From%
\begin{equation}
\left\vert 2\epsilon \ell ^{2}\left( \frac{\overline{p}^{0}}{\overline{\Im }}%
\right) ^{2}\right\vert \leq 3\times 10^{-6}  \label{2.11a}
\end{equation}%
with $\overline{p}^{0}=17$ GeV and the eigenvalue of the operator $\overline{%
\Im }$, in the right hand side of the Heisenberg algebra, set to $\overline{%
\Im }=1$\footnote{%
In the framework of the representations of some subalgebras \cite{VilelaPLA2}
of (\ref{A1}), an explicit representation of $\overline{\Im }$ as $\overline{%
\Im }=\left( 1+\ell ^{2}\overline{p}^{2}\right) ^{1/2}$ is possible. However
this does not change the $O\left( \ell ^{2}\right) $ wave packet speed
correction.}, it follows\footnote{%
Notice that the correction due to a neutrino mass $\sim 2$ eV is smaller, of
order $10^{-19}$}%
\begin{equation}
\ell \leq 1.4\times 10^{-18}\text{cm}  \label{2.11b}
\end{equation}%
or, equivalently, for the elementary time%
\begin{equation}
\tau \leq 0.5\times 10^{-28}\text{sec}  \label{2.11c}
\end{equation}

From the MINOS \cite{MINOS}\ data, with neutrino spectrum peaked at $%
\overline{p}^{0}=3$ GeV%
\begin{equation}
\left\vert \frac{v-c}{c}\right\vert =\left( 5.1\pm 2.9\right) \times 10^{-5}
\label{2.12}
\end{equation}%
\begin{equation}
\left\vert 2\epsilon \ell ^{2}\left( \frac{\overline{p}^{0}}{\overline{\Im }}%
\right) ^{2}\right\vert \leq 5\times 10^{-5}  \label{2.12a}
\end{equation}%
one obtains%
\begin{equation}
\ell \leq 3.3\times 10^{-17}\text{cm;\hspace{2cm}}\tau \leq 10^{-27}\text{sec%
}  \label{2.12b}
\end{equation}

Assuming a delay of at most a couple of hours between the neutrino and the
visible light outbursts from the SN1987A supernova several authors \cite{SN1987A} \cite{SN2} \cite{SN3} have estimated 
\begin{equation}
\left\vert \frac{v-c}{c}\right\vert <2\times 10^{-9}  \label{2.13}
\end{equation}%
which with $\overline{p}^{0}\approx 10$ MeV would lead to%
\begin{equation}
\ell <6\times 10^{-17}\text{cm;\hspace{2cm}}\tau <2\times 10^{-27}\text{sec}
\label{2.13a}
\end{equation}

One sees that all this data is compatible with a value $\ell \lesssim
10^{-18}$cm or $\tau \lesssim 0.3\times 10^{-28}$sec. Using this value one
also sees that the effect is extremely small for visible light. For example
with $\overline{p}^{0}=3$ eV and $\ell =10^{-18}$ cm one obtains%
\begin{equation*}
\left\vert \frac{v-c}{c}\right\vert <4.6\times 10^{-26}
\end{equation*}

These are results for elementary states. For slow macroscopic matter instead
of (\ref{2.2}) the state is%
\begin{equation}
\psi _{0}\left( P\right) =\int \left\vert k_{1},k_{2},\cdots
,k_{N}\right\rangle f_{P}\left( k_{1},k_{2},\cdots ,k_{N}\right)
dk_{1}dk_{2}\cdots dk_{N}  \label{2.14}
\end{equation}%
Whenever the coupling energy of the elementary constituents of the
macroscopic body is much smaller than their rest masses one may factorize
the time shift operator%
\begin{equation}
e^{-it\Upsilon }\left\vert k_{1},k_{2},\cdots ,k_{N}\right\rangle
=\left\vert e^{-it\Upsilon _{1}}k_{1},e^{-it\Upsilon _{2}}k_{2},\cdots
,e^{-it\Upsilon _{N}}k_{N}\right\rangle  \label{2.14a}
\end{equation}%
Therefore for a nonrelativistic body $\overline{p}^{0}\simeq m_{p}$ (the
proton mass $m_{p}=938$ MeV) leads, with $\ell =10^{-18}$ cm, to a speed
correction%
\begin{equation}
\left\vert \frac{v-\frac{\overline{p}}{\overline{p}^{0}}}{\frac{\overline{p}%
}{\overline{p}^{0}}}\right\vert =0.452\times 10^{-8}  \label{2.14b}
\end{equation}%
It does not sound like much, however, for a nominal velocity $\frac{%
\overline{p}}{\overline{p}^{0}}=10$ Km/sec it would lead after one year to a
deviation of $1.4$ Km.

All the above bounds are much larger than the Planck's time scale and
improving them seems in reach of present experimental techniques. In
particular, it would be interesting to refine the neutrino wave packet speed
measurements, preferably with a larger baseline.

Presumably the best way to test the speed corrections arriving from
noncommutativity would be to consider phenomena involving cosmological
distances. This is also the point of view of many authors when looking for
light velocity modifications as a probe of Lorentz invariance violation
(LIV) (\cite{AmelinoGRB} \cite{Xu1} \cite{Ellis2} and references therein).
In particular special attention has been devoted to gamma ray bursts (GRB).
Notice however that in the present paper no LIV is implied, it is the
noncommutativity that impacts the group velocity of massless particle wave
packets. In any case the LIV-estimates of these authors may in some cases be
carried over to the noncommutativity framework and I will comment on that
later.

As will be seen, the calculation of cosmological distances (angular diameter
and luminosity distance) is affected by the energy-dependent wave packet
speed corrections.

One uses the Robertson-Walker metric%
\begin{equation}
\left( ds\right) ^{2}=\left( dt\right) ^{2}-a^{2}\left( t\right) \left\{ 
\frac{\left( dr\right) ^{2}}{1-Kr^{2}}+r^{2}\left( \left( d\theta \right)
^{2}+\sin ^{2}\theta \left( d\phi \right) ^{2}\right) \right\}  \label{COS1}
\end{equation}%
($c=\hslash =1$).

For a massless wave packet with central energy $E$ moving radially at speed $%
v\left( E\right) $%
\begin{equation}
v\left( E\left( t\right) \right) \frac{dt}{a\left( t\right) }=\frac{dr}{%
\sqrt{1-Kr^{2}}}  \label{COS2}
\end{equation}%
with, in leading $\ell ^{2}$ order%
\begin{equation}
v\left( E\left( t\right) \right) =1-2\epsilon \ell ^{2}E^{2}\left( t\right)
=1-8\pi ^{2}\epsilon \ell ^{2}\frac{1}{\lambda ^{2}\left( t\right) }
\label{COS3}
\end{equation}%
$\lambda $ being the wavelength. Considering now two crests in the central
frequency of the packet, using (\ref{COS2})%
\begin{equation*}
\frac{v\left( E_{0}\right) }{a\left( t_{0}\right) }\lambda _{0}=\frac{%
v\left( E_{e}\right) }{a\left( t_{e}\right) }\lambda _{e}
\end{equation*}
which in leading $\ell ^{2}$ order is%
\begin{equation}
\frac{\lambda _{e}}{\lambda _{0}}\left\{ 1-8\pi ^{2}\epsilon \ell ^{2}\left( 
\frac{1}{\lambda _{e}^{2}}-\frac{1}{\lambda _{0}^{2}}\right) \right\} =\frac{%
a\left( t_{e}\right) }{a\left( t_{0}\right) }  \label{COS4}
\end{equation}%
$\lambda _{e}$ being the emitted wavelength at time $t_{e}$ and $\lambda
_{0} $ the received one at time $t_{0}$. Defining $1+z=\frac{\lambda _{0}}{%
\lambda _{e}}$%
\begin{equation}
\frac{a\left( t_{0}\right) }{a\left( t_{e}\right) }=\frac{1+z}{\Gamma \left(
\lambda _{0},z\right) }  \label{COS5}
\end{equation}%
with%
\begin{equation}
\Gamma \left( \lambda _{0},z\right) =1-8\pi ^{2}\epsilon \frac{\ell ^{2}}{%
\lambda _{0}^{2}}z\left( z+2\right)  \label{COS6}
\end{equation}%
Therefore the relation between the ratio $\frac{a\left( t_{0}\right) }{%
a\left( t_{e}\right) }$ and the redshift $z$ depends on the frequency that
is being observed, that is, when using integration over redshift, to obtain
the propagation time, one should take into account the wavelength for which
the redshift is being measured. From (\ref{COS5}) one obtains%
\begin{equation}
dt=\frac{1}{H\left( t\right) }\left( \frac{d\log \Gamma \left( \lambda
_{0},z\right) }{dz}-\frac{1}{1+z}\right) dz  \label{COS7}
\end{equation}%
$H\left( t\right) $ being the Hubble parameter,%
\begin{equation}
H\left( t\right) =\frac{\overset{\bullet }{a}\left( t\right) }{a\left(
t\right) }  \label{COS8}
\end{equation}

The Friedmann equation becomes%
\begin{eqnarray}
\frac{H\left( t\right) }{H_{0}} &=&\sqrt{\sum_{i}\Omega _{i,0}\left( \frac{%
1+z}{\Gamma \left( \lambda _{0},z\right) }\right) ^{3}+\Omega _{rad,0}\left( 
\frac{1+z}{\Gamma \left( \lambda _{0},z\right) }\right) ^{4}+\Omega
_{K,0}\left( \frac{1+z}{\Gamma \left( \lambda _{0},z\right) }\right)
^{2}+\Omega _{\Lambda ,0}}  \notag \\
&=&\sqrt{E\left( \lambda _{0},z\right) }  \label{COS9}
\end{eqnarray}%
the $\Omega $ constants related, respectively, to matter, radiation,
curvature and vacuum energy. The dependence on $\lambda _{0}$ means that the
redshift $z$ is computed from the received light at $\lambda _{0}$
wavelength.

The dependence on $\lambda _{0}$ would also have an impact on estimates of
the age of the universe%
\begin{equation}
t_{0}=\frac{1}{H_{0}}\int_{0}^{\infty }\frac{dz}{\sqrt{E\left( \lambda
_{0},z\right) }}\left( \frac{1}{1+z}-\frac{d\log \Gamma \left( \lambda
_{0},z\right) }{dz}\right)  \label{COS10}
\end{equation}%
For the angular diameter $d_{A}$ and luminosity $d_{L}$ distances one has%
\begin{equation}
d_{A}=\frac{\Gamma \left( \lambda _{0},z\right) }{1+z}F_{K}\left( \frac{1}{%
H_{0}}\int_{0}^{z}\frac{\left( 1+z\right) \left( \frac{1}{1+z}-\frac{d\log
\Gamma \left( \lambda _{0},z\right) }{dz}\right) \left( 1-8\pi ^{2}\epsilon 
\frac{\ell ^{2}\left( 1+z\right) ^{2}}{\lambda _{0}^{2}}\right) }{\Gamma
\left( \lambda _{0},z\right) \sqrt{E\left( \lambda _{0},z\right) }}dz\right)
\label{COS12}
\end{equation}%
\begin{equation}
d_{L}=\frac{\left( 1+z\right) ^{2}}{\Gamma ^{2}\left( \lambda _{0},z\right) }%
d_{A}  \label{COS13}
\end{equation}%
with $F_{K}=\left\{ 
\begin{array}{c}
\sin \\ 
1 \\ 
\sinh%
\end{array}%
\right. $ for $K=\left\{ 
\begin{array}{c}
1 \\ 
0 \\ 
-1%
\end{array}%
\right. $.

With these results some experimental information might be obtained from
cosmological data. As an example consider the spectral lags \cite{Norris1} 
\cite{Norris2} \cite{Ukwatta1} \cite{Ukwatta2} \cite{Shao} in gamma ray
bursts (GRB). The spectral lag is defined as the difference in time of
arrival of high and low energy photons. It is considered positive when the
high energy photons arrive earlier than the low energy ones. The spectral
lags being associated to the spectral evolution during the prompt GBR phase,
one expects different source types to have different intrinsic lags at the
source. In addition, due to the complex nature of the gamma-ray peak
structure, the spectral lags, obtained from delayed correlation
measurements, have large error bars. Nevertheless they allow access to time
scales not achievable in the labs and it might be worthwhile to test whether
the lags are also affected by energy-dependent propagation effects. A few
simple hypothesis will be made about the relation between the lag in the
production of gamma rays at the source and their observation at earth. Let
us consider two gamma pulses at different energies $E_{1}$ and $E_{2}$ $%
\left( E_{2}>E_{1}\right) $ produced with an intrinsic lag $\alpha ^{\left(
a\right) }$ at the source $a$.

If $T_{a}^{(1)}$ and $T_{a}^{(2)}$ are their propagation times from the
source $a$ to earth, the spectral lag would be%
\begin{equation}
\Delta t_{a}=T_{a}^{(1)}-T_{a}^{(2)}+\alpha ^{\left( a\right) }  \label{2.15}
\end{equation}%
From (\ref{COS7}) and (\ref{COS9})%
\begin{equation}
T^{(i)}\left( \lambda _{0}^{(i)},z\right) =\frac{1}{H_{0}}%
\int_{0}^{z}dz^{\prime }\frac{1}{\sqrt{E\left( z^{\prime }\right) }}\left\{ 
\frac{1}{1+z^{\prime }}-\frac{d}{dz^{^{\prime }}}\log \Gamma \left( \lambda
_{0}^{(i)},z\right) \right\}  \label{2A.1}
\end{equation}%
with $\lambda _{0}^{(i)}$ the wavelengths as observed at earth and $E\left(
z\right) $ defined in (\ref{COS9}). Adopting the nowadays consensus
cosmology $\Omega _{m,0}=0.3$, $\Omega _{\Lambda ,0}=0.7$, $\Omega
_{k,0}=\Omega _{rad,0}=0$, $K=0$, $T^{(i)}\left( \lambda _{0},z\right) $
becomes in leading $\frac{\ell ^{2}}{\lambda _{0}^{2}}$ order%
\begin{equation}
T^{(i)}\left( \lambda _{0}^{(i)},z\right) \simeq \frac{I_{1}\left( z\right) 
}{H_{0}}+\frac{4\pi ^{2}\epsilon \ell ^{2}}{H_{0}\lambda _{0}^{(i)2}}%
I_{2}\left( z\right)  \label{2A.2}
\end{equation}%
the integrals $I_{1}$ and $I_{2}$ being%
\begin{eqnarray}
I_{1}\left( z\right) &=&\int_{0}^{z}\frac{dz^{\prime }}{1+z^{\prime }}\frac{1%
}{\left( \Omega _{m,0}\left( 1+z^{\prime }\right) ^{3}+\Omega _{\Lambda
,0}\right) ^{\frac{1}{2}}}  \notag \\
I_{2}\left( z\right) &=&\int_{0}^{z}dz^{\prime }\frac{\Omega _{m,0}\left(
1+z^{\prime }\right) ^{2}\left( z^{\prime 2}+2z^{\prime }+4\right) +4\left(
1+z^{\prime }\right) \Omega _{\Lambda ,0}}{\left( \Omega _{m,0}\left(
1+z^{\prime }\right) ^{3}+\Omega _{\Lambda ,0}\right) ^{\frac{3}{2}}}
\label{2A.3}
\end{eqnarray}

From (\ref{2.15}) and (\ref{2A.2}) one sees that the lags are linear on $%
I_{2}\left( z\right) $, 
\begin{equation*}
\Delta t_{a}=\frac{4\pi ^{2}\epsilon \ell ^{2}}{H_{0}}I_{2}\left( z\right)
\left( \frac{1}{\lambda _{0}^{(1)2}}-\frac{1}{\lambda _{0}^{(2)2}}\right)
+\alpha _{a}^{(0)}
\end{equation*}%
for wavelengths at earth or on $\frac{I_{2}\left( z\right) }{\left(
1+z\right) ^{2}}$ for energies at the source%
\begin{eqnarray}
\Delta t_{a} &=&\frac{4\pi ^{2}\epsilon \ell ^{2}}{H_{0}}\frac{I_{2}\left(
z\right) }{\left( 1+z\right) ^{2}}\left( \frac{1}{\lambda _{e}^{(1)2}}-\frac{%
1}{\lambda _{e}^{(2)2}}\right) +\alpha _{a}^{(0)}  \notag \\
&=&\frac{\epsilon \ell ^{2}}{H_{0}}\frac{I_{2}\left( z\right) }{\left(
1+z\right) ^{2}}\left( E_{e}^{(1)2}-E_{e}^{(2)2}\right) +\alpha _{a}^{(0)}
\label{2.16}
\end{eqnarray}%
Thus, for fixed $\left( E_{e}^{(1)2}-E_{e}^{(2)2}\right) $ one may expect
the data to be fitted by a few parallel lines, each one corresponding to a
particular type of lag mechanism at the source. This analysis is similar to
what has been done by other authors (see for example \cite{Xu1} \cite{Zhang1}%
) in the context of \ searches for LIV.

Let $H_{0}=70$ Km s$^{-1}$, $\Omega _{m,0}=0.3$, $\Omega _{\Lambda ,0}=0.7$
and to test the hypothesis, use the Swift BAT data on reference \cite%
{Ukwatta2} for spectral lags of the source-frame bands $100-150$ KeV and $%
200-250$ Kev ($E_{0}^{(1)}=\frac{125}{1+z}$, $E_{0}^{(2)}=\frac{225}{1+z}$),
selecting the $24$ bursts for which the lags were computed with significance 
$1\sigma $ or greater. The following table lists the correspondence of the
numbers in the plot with the burst code.%
\begin{equation*}
\begin{tabular}{|l|l|l|l|}
\hline
1 & GRB050401 & 13 & GRB080413B \\ \hline
2 & GRB050922C & 14 & GRB080605 \\ \hline
3 & GRB051111 & 15 & GRB080916A \\ \hline
4 & GRB060210 & 16 & GRB081222 \\ \hline
5 & GRB061007 & 17 & GRB090618 \\ \hline
6 & GRB061121 & 18 & GRB090715B \\ \hline
7 & GRB071010B & 19 & GRB090926B \\ \hline
8 & GRB071020 & 20 & GRB091024 \\ \hline
9 & GRB080319B & 21 & GRB091208B \\ \hline
10 & GRB080319C & 22 & GRB100621A \\ \hline
11 & GRB080411 & 23 & GRB100814A \\ \hline
12 & GRB080413A & 24 & GRB100906A \\ \hline
\end{tabular}%
\end{equation*}%
Comparison of the data with Eq.(\ref{2.16}) is performed by minimizing in $%
\beta $ and $\overrightarrow{\alpha }$ the function%
\begin{equation*}
f\left( \beta ,\overrightarrow{\alpha }\right) =\sum_{i}\min_{%
\overrightarrow{\alpha }}\left\{ y_{i}-\left( \beta x_{i}+\overrightarrow{%
\alpha }\right) \right\} ^{2}
\end{equation*}%
for several dimensions of the vector $\overrightarrow{\alpha }$ (the vector
of intrinsic lags). Here the variables $y_{i}$ and $x_{i}$ are respectively
the observed lags $\Delta t_{a}$ and $\frac{I_{2}\left( z\right) }{\left(
1+z\right) ^{2}}$. As the dimension of $\overrightarrow{\alpha }$ (the
number of different lag types at the sources) increases, the fitting error,
defined as%
\begin{equation*}
er=\frac{f\left( \beta ,\overrightarrow{\alpha }\right) }{\sum_{i}y_{i}^{2}}
\end{equation*}%
decreases. The fitting accuracy improves appreciably until dimension of $%
\overrightarrow{\alpha }^{(0)}$ equal to $3$, but not much afterwards.

\begin{figure}[htb]
    \centering
    \includegraphics[width=0.5\textwidth]{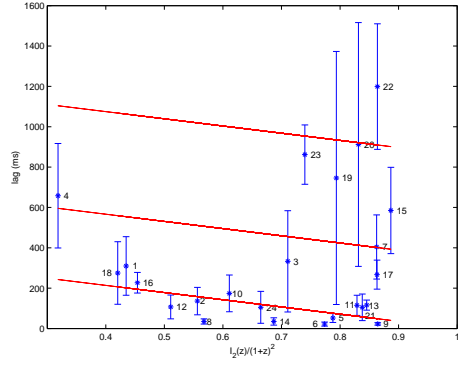}
    \caption{Lags versus $\frac{I_{2}\left(
z\right) }{\left( 1+z\right) ^{2}}$ for 24 GRB's with 1$\protect\sigma $
significance or greater (data from \protect\cite{Ukwatta2})}
    \label{fit_I2}
\end{figure}

\begin{figure}[htb]
    \centering
    \includegraphics[width=0.5\textwidth]{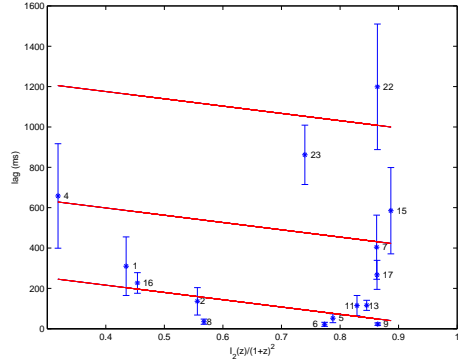}
    \caption{Lags versus $\frac{I_{2}\left( z\right) }{\left(
1+z\right) ^{2}}$ for 15 GRB's with significance 2$\protect\sigma $ or
greater (data from \protect\cite{Ukwatta2})}
    \label{fit_I2_2s}
\end{figure}

The figure (\ref%
{fit_I2}) shows the data points and the fitting lines for $\dim 
\overrightarrow{\alpha }^{(0)}=3$. The error is $er=0.05$. The slope $\beta $
is $\simeq $ $-360$ corresponding, with $\left( E_{1}^{2}-E_{2}^{2}\right)
=35000$ KeV$^{2}$ to $\ell \simeq 0.95\times 10^{-19}$cm and $\epsilon =+1$
( $\tau =3\times 10^{-30}$s). Notice that $I_{2}\left( z\right) $ grows with 
$z$ but not $\frac{I_{2}\left( z\right) }{\left( 1+z\right) ^{2}}$. As shown
in Fig.(\ref{fit_I2_2s}) the result is quite similar when one restricts to
the GBR's with significance 2$\sigma $ or greater.

Notice that $\epsilon =+1$ corresponds to higher energy pulses travelling
slower than lower energy ones.

\begin{figure}[htb]
    \centering
    \includegraphics[width=0.5\textwidth]{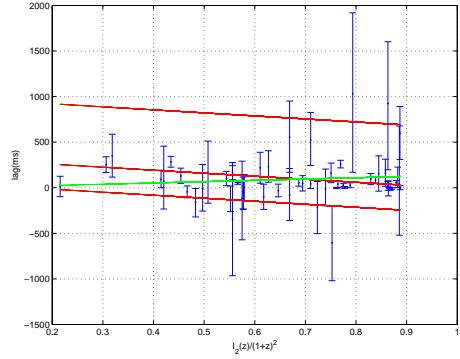}
    \caption{Lags versus $\frac{I_{2}\left( z\right) }{\left( 1+z\right) ^{2}}$ for the 57 GRB's in 
\protect\cite{Bernardini} and fitting lines with 3 (red) or one intersect (green)}
    \label{Bernardini}
\end{figure}

A larger set of GRB data with known redshifts is studied in \cite{Bernardini}%
. The main difference from the analysis in \cite{Ukwatta2} is the use of an
asymmetric Gaussian model for the cross-correlation function to compute the
spectral lags. Otherwise the source frame energy bands ($100-150$ and $%
200-250$ KeV) are the same as in \cite{Ukwatta2}. The same fitting technique
as before was here applied to the 57 GRB's in \cite{Bernardini} with the
result shown in Fig.\ref{Bernardini}. For 3 intersects (dimension of $%
\overrightarrow{\alpha }^{(0)}=3$) the slope that is obtained is $\beta
\simeq -330$ ($er=0.18$) corresponding to $\ell \simeq 0.9\times 10^{-19}$%
cm, $\epsilon =+1$, a result consistent with the one obtained before. Notice
however that if instead of dimension of $\overrightarrow{\alpha }^{(0)}=3$
one assumes dimension of $\overrightarrow{\alpha }^{(0)}=1$ one obtains a
worse fit ($er=0.87$) and a quite different result, that is $\ell \simeq 0$,
the dash-dotted green line in Fig.\ref{Bernardini}. This is essentially what
has been done in \cite{Wei} with these authors concluding that there is no
evidence for LIV. However that hypothesis (dimension of $\overrightarrow{%
\alpha }^{(0)}=1$) assumes that all the intrinsic lags at the source are the
same.

\begin{figure}[htb]
    \centering
    \includegraphics[width=0.5\textwidth]{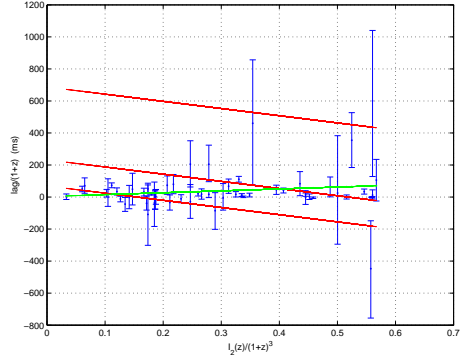}
    \caption{Lags/(1+z) versus $\frac{I_{2}\left( z\right) }{\left( 1+z\right) ^{3}}$ for
the 57 GRB's in \protect\cite{Bernardini} and fitting lines with 3 (red) or
one intersect (green)}
    \label{Bernardini_2}
\end{figure}

In Eq.(\ref{2.16}) the $\alpha _{a}^{(0)}$ line intersects represent several
classes of intrinsic lags as seen at earth. It might be better to use the
intrinsic lags $\alpha _{a}^{(e)}$ at the source, that is replace $\alpha
_{a}^{(0)}$ by $\alpha _{a}^{(0)}=\alpha _{a}^{(e)}\left( 1+z\right) $. Then
one has%
\begin{equation}
\frac{\Delta t_{a}}{1+z}=\frac{\epsilon \ell ^{2}}{H_{0}}\frac{I_{2}\left(
z\right) }{\left( 1+z\right) ^{3}}\left( E_{e}^{(1)2}-E_{e}^{(2)2}\right)
+\alpha _{a}^{(e)}  \label{2.17}
\end{equation}%
With this equation and the data in \cite{Bernardini} one obtains the results
shown in Fig.\ref{Bernardini_2}

For dimension
of $\overrightarrow{\alpha }^{(e)}=3$ the slope is $\beta =-448$ ($er=0.2$)
corresponding with $\left( E_{e}^{(1)2}-E_{e}^{(2)2}\right) =3.5\times
10^{4} $ KeV$^{2}$ to $\ell \simeq 1.06\times 10^{-19}$cm. And, as before, a
very different result is obtained for dimension of $\overrightarrow{\alpha }%
^{(e)}=1$ ($er=0.9$), the dash-dotted line in Fig.\ref{Bernardini_2}.

\begin{figure}[htb]
    \centering
    \includegraphics[width=0.5\textwidth]{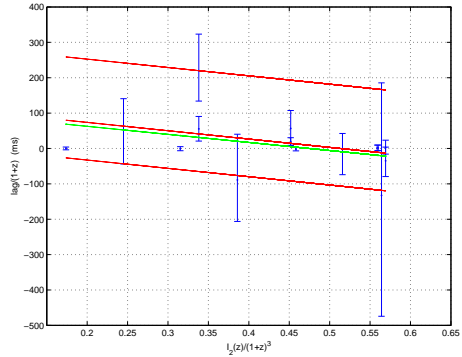}
    \caption{Lags/(1+z) versus $\frac{I_{2}\left( z\right) }{\left( 1+z\right) ^{3}}$ for
the 15 short GRB's in \protect\cite{Bernardini2} and fitting lines with 3
(red) or one intersect (green)}
    \label{Bernardini_short}
\end{figure}

For short GRB pulses intrinsic lags are in general considered smaller that
those of long GRB pulses. Looking for eventual Lorentz invariance violation
(LIV), the authors in \cite{Bernardini2} have analyzed 15 short pulses (on
the energy bands 50-100 and 150-200 KeV) concluding that there is no evidence%
\footnote{%
Actually the authors conclusion is that the quantum gravity scale $%
E_{QG}\gtrsim 1.5\times 10^{16}$GeV, which would correspond to a scale $\ell
\lesssim 1.3\times 10^{-30}$cm.} for energy dependence of the light
propagation speed. Here the same analyzing technique as described above has
been applied to the same data with a different conclusion, as shown in Fig.%
\ref{Bernardini_short}

The
figure shows the fitting of the data assuming either dimension of $%
\overrightarrow{\alpha }^{(e)}=3$ (red continuous lines) of dimension of $%
\overrightarrow{\alpha }^{(e)}=1$ (dash-dotted green line). The slope is $%
\beta \simeq -230$ corresponding with $\left(
E_{e}^{(1)2}-E_{e}^{(2)2}\right) =2.5\times 10^{4}$ KeV$^{2}$ to $\ell
\simeq 0.9\times 10^{-19}$cm. Notice that in this case the slope obtained
with one or three intersects is essentially the same, suggesting that for
this set of pulses the intrinsic lags are identical. The difference to the
conclusions of the authors in \cite{Bernardini2} are not, of course, due to
any mistake of these authors but to the fact that they plot the data with
respect to a $K\left( z\right) $ function, whereas here, according to the
calculations above, the $z-$dependence is coded by the $I_{2}\left( z\right) 
$ function (Eq.\ref{2A.3}).

Of course, all these results, as well as the searches for LIV (see for
example \cite{Ellis2} and references therein), can only be taken as
indicative or as establishing an upper bound on $\tau $ because of the large
uncertainties on the calculation of the spectral lags, on the statistics of
the GRB pulses and even more on the intrinsic spectral lags $\alpha _{a}$.
However, if correct, they have some implications concerning the observation
of neutrino emissions from the GRB sources and also on the SN1987A
observations. From the SN1987A supernova, neutrinos were observed in the
range from $7.5$ to $40$ Mev\cite{SN1987A} \cite{SN2}. Using Eq.(\ref{2.9})
to obtain the propagation time difference over $168000$ light years, between
visible light and neutrino packets of $10$ and $40$ Mev, with $\ell
=10^{-19} $ cm, one obtains respectively $1.1\times 10^{-3}$ and $1.8\times
10^{-2}$ seconds. Clearly this does not change the estimate in (\ref{2.13}).
However for GRB's at cosmological distances the situation is different. From
(\ref{COS7}) $\ell =10^{-19}$ cm and $\epsilon =+1$, neutrinos of energy $40$
Mev, would take $27$ hours more than visible light to reach earth from a
source at $z=2$ redshift and $1.7$ more hours from a source at redshift $z=1$%
. For $10$ MeV neutrinos the result would be $1.7$ and $1$ hour. For $\ell
=10^{-18}$ cm these numbers would be multiplied by $100$ and also grow
quadratically with the energy.

Recently a very high energy neutrino was observed from the direction of
active galactic nuclei at cosmological distance \cite{Blazar1} \cite{Blazar2}%
. If $\ell $ is in the range discussed above, the conclusion is that it
could only have originated from a much earlier event, not a recent flare of
gamma activity. Alternatively if by some means its origin is proved to be
coincident with recently observed gamma flares, that would mean that $\ell $
is much smaller than suggested here (that is, $\ell \preceq 10^{-24}$cm).
Notice however that dedicated searches \cite{Abbasi} \cite{Adrian} \cite%
{Aartsen} for neutrinos in close coincidence with GRB bursts found no or
scarce evidence for them.

\begin{figure}[htb]
    \centering
    \includegraphics[width=0.5\textwidth]{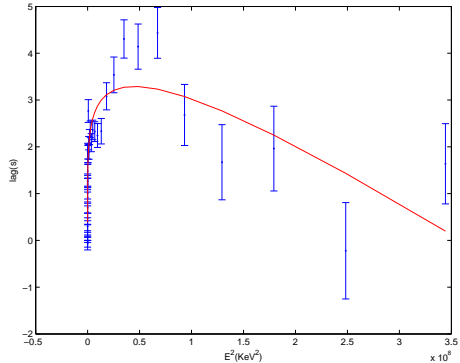}
    \caption{Fitting the GRB160625B data to
Eq.(\protect\ref{COS20})}
    \label{grb160625b_bad}
\end{figure}

Wei et al. \cite{Wei1} \cite{Wei2} analyzed a burst GRB160625B with
unusually high photon statistics and a steep decline from positive lags to
smaller ones with increasing photon energy in the range 8-20 MeV. They have
fitted the spectral lag data using a power law for the intrinsic lag and a
linear or quadratic term corresponding to the LIV correction. Here the same
data has been analyzed using also a power law for the intrinsic lag together
with the noncommutativity correction, namely%
\begin{equation}
lag=\alpha E^{\beta }-\frac{\epsilon \ell ^{2}}{H_{0}}I_{2}\left(
1.41\right) E^{2}  \label{COS20}
\end{equation}%
The least squares result is shown in Fig.(\ref{grb160625b_bad}). One sees
that the fitting accuracy is rather poor, what is even more apparent using a
linear $E^{2}$ axis than in the log-log plot used in \cite{Wei1}. Actually
the small statistical significance of the fitting using an equation of the
type of Eq.(\ref{COS20}) had already been pointed out in \cite{Ganguly}.%

\begin{figure}[htb]
    \centering
    \includegraphics[width=0.5\textwidth]{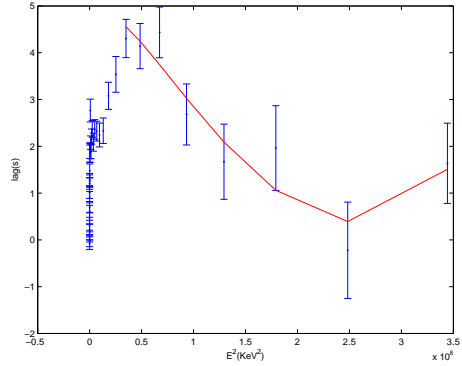}
    \caption{Fitting the GRB160625B data between 5 and 20 GeV by Eq.(\protect\ref{COS21})}
    \label{grb160625b_31_38}
\end{figure}

In fact given the probable multiple shock mechanism of the GRB's generation
is not likely that a continuous power dependence of the intrinsic lag be a
good hypothesis. It seems better to concentrate on the high energy tail of
the data and try the equation%
\begin{equation}
lag=\alpha E^{0.18+\beta E^{2}}-\frac{\epsilon \ell ^{2}}{H_{0}}I_{2}\left(
1.41\right) E^{2}  \label{COS21}
\end{equation}%
This is used to fit the data between $5-20$ GeV the result being shown in
Fig.(\ref{grb160625b_31_38}).

The
minimizing parameters are $\alpha =0.29,\beta =1.538\times 10^{-10}$ and%
\begin{equation}
\frac{\epsilon \ell ^{2}}{H_{0}}I_{2}\left( 1.41\right) =7.794\times 10^{-8}
\label{COS22}
\end{equation}%
$0.18$ in the exponent being the value obtained in the fitting to Eq.(\ref%
{COS20}). With $I_{2}\left( 1.41\right) =4.3445$ one obtains from (\ref%
{COS22}) $\ell =1.79\times 10^{-21}$ cm or $\tau =0.597\times 10^{-31}$ s.
This is two orders of magnitude smaller than obtained before, but there is
small significance of a result obtained with a single burst, compounded with
the small quantitative knowledge that still exists about the intrinsic lags
at the source.

Finally, from (\ref{COS13}) one may also estimate the impact of an energy
dependent propagation speed on the calculation of the Hubble constant from
observations at cosmological distances. \ Given the luminosity $L$ and the
observed flux $F_{o}$ from a standard candle, the luminosity distance $d_{L}$
is%
\begin{equation}
d_{L}^{2}=\frac{L}{4\pi F_{o}}  \label{COS14}
\end{equation}%
On the other hand from (\ref{COS12}) and (\ref{COS13})%
\begin{equation}
d_{L}=\frac{1+z}{\Gamma \left( \lambda _{0},z\right) }I\left( z\right)
\label{COS15}
\end{equation}%
with%
\begin{equation}
I\left( z\right) =\frac{1}{H_{0}}\int_{0}^{z}\frac{\left( 1+z\right) \left( 
\frac{1}{1+z}-\frac{d\log \Gamma \left( \lambda _{0},z\right) }{dz}\right)
\left( 1-8\pi ^{2}\frac{\ell ^{2}}{\lambda _{0}^{2}}\frac{\left( 1+z\right)
^{2}}{\Gamma ^{2}\left( \lambda _{0},z\right) }\right) }{\Gamma \left(
\lambda _{0},z\right) \sqrt{E\left( \lambda _{0},z\right) }}dz  \label{COS16}
\end{equation}%
Therefore given $d_{L}$ from (\ref{COS14}), $H_{0}$ is obtained from%
\begin{equation*}
H_{0}=\frac{\left( 1+z\right) I_{0}\left( z\right) }{d_{L}}+\frac{\pi
^{2}\epsilon \ell ^{2}}{\lambda _{0}^{2}}\left( 1+z\right) \left\{
I_{3}\left( z\right) +8z\left( z+2I_{0}\left( z\right) \right) \right\}
\end{equation*}%
with%
\begin{eqnarray*}
I_{0}\left( z\right) &=&\int_{0}^{z}\frac{dz^{\prime }}{\left( \Omega
_{m,0}\left( 1+z^{\prime }\right) ^{3}+\Omega _{\Lambda ,0}\right) ^{\frac{1%
}{2}}} \\
I_{3}\left( z\right) &=&\int_{0}^{z}\left\{ \frac{2z^{\prime 2}+8z^{\prime
}+3}{\left( \Omega _{m,0}\left( 1+z^{\prime }\right) ^{3}+\Omega _{\Lambda
,0}\right) ^{\frac{1}{2}}}-\frac{12z^{\prime }\left( z^{\prime }+2\right)
\Omega _{m,0})\left( 1+z^{\prime }\right) ^{3}}{\left( \Omega _{m,0}\left(
1+z^{\prime }\right) ^{3}+\Omega _{\Lambda ,0}\right) ^{\frac{3}{2}}}%
\right\} dz^{\prime }
\end{eqnarray*}%
Then the correction to the $H_{0}$ calculation is%
\begin{equation*}
\frac{H_{0}\left( \ell \neq 0\right) }{H_{0}\left( \ell =0\right) }=1+\frac{%
\pi \epsilon \ell ^{2}}{\lambda _{0}^{2}}\left\{ \frac{I_{3}\left( z\right) 
}{I_{0}\left( z\right) }+8z\left( z+2\right) \right\}
\end{equation*}%
However, for example for $z=0.5$ this would be $1+17.3\frac{\pi \epsilon
\ell ^{2}}{\lambda _{0}^{2}}$, which for visible light ($\lambda
_{0}=3.9-7\times 10^{-9}$cm) would be too small to be of any importance.
Hence this correction should not be relevant to the present $H_{0}$ tension
problem.

\subsection{Corrections to the Kepler problem}

By the Kepler problem one means motion of a body under the central $\frac{1}{%
r}$ potential. In reality an additional inverse cubic term should be added
to account for the general relativity corrections. Here only the
modifications to the $\frac{1}{r}$ term arising from noncommutativity will
be considered (in first $\ell ^{2}$ order). First one considers the
corrections to the classical Hamiltonian%
\begin{equation}
H=\frac{p^{2}}{2m}+\frac{G}{r}  \label{Ke1}
\end{equation}%
Using the representation (\ref{2.6}) and taking expectation values in a
basis $e^{ir\cdot \frac{\overline{p}}{\overline{\Im }}}$ ($r$ here is a
c-number, not an operator), 
\begin{eqnarray*}
\sum_{i}\left( e^{ir\cdot \frac{\overline{p}}{\overline{\Im }}%
},x^{i}x^{i}e^{ir\cdot \frac{\overline{p}}{\overline{\Im }}}\right) 
&=&\sum_{i}\left( \left( r_{i}+\epsilon \ell ^{2}p^{i}\frac{r\cdot \overline{%
p}}{\overline{\Im }^{2}}\right) e^{ir\cdot \frac{\overline{p}}{\overline{\Im 
}}},\left( r_{i}+\epsilon \ell ^{2}p^{i}\frac{r\cdot \overline{p}}{\overline{%
\Im }^{2}}\right) e^{ir\cdot \frac{\overline{p}}{\overline{\Im }}}\right)  \\
&=&\left\vert r\right\vert ^{2}\left( 1-2\epsilon \ell ^{2}\frac{\left(
r\cdot \overline{p}\right) ^{2}}{\overline{\Im }^{2}}\right) +O\left( \ell
^{4}\right) 
\end{eqnarray*}
one sees that the $O\left( \ell ^{2}\right) $ corrections to the classical
Hamiltonian (\ref{Ke1}) amount to the replacement%
\begin{equation}
\frac{G}{r}\rightarrow \frac{G}{r}\left( 1+\epsilon \ell ^{2}\left( \widehat{%
r}\cdot \frac{\overline{p}}{\overline{\Im }}\right) ^{2}+O\left( \ell
^{4}\right) \right)   \label{Ke3}
\end{equation}%
$\widehat{r}=\frac{r}{\left\vert r\right\vert }$. One obtains a positive or
negative correction (depending on $\epsilon $) of the coupling constant. For
classical bound quasi-circular orbits $\left( \widehat{r}\cdot \frac{%
\overline{p}}{\overline{\Im }}\right) $ is very small, therefore any
detectable corrections to the classical motion could only be expected for
flyby orbits.

With the estimate%
\begin{equation*}
\int_{\text{collision}}\left( \widehat{r}\cdot \overline{p}\right)
^{2}=\int_{0}^{\pi }\left\vert \overline{p}\right\vert ^{2}\cos \theta
d\theta =\left\vert \overline{p}\right\vert ^{2}\frac{\pi }{2}
\end{equation*}%
the approximate correction to the coupling constant would be%
\begin{equation*}
G\rightarrow G\left( 1+\epsilon \ell ^{2}\left\vert \overline{p}\right\vert
^{2}\frac{\pi }{2}\right) 
\end{equation*}%
which for a macroscopic speed $15$ Km/sec, the proton mass and $\ell
=10^{-19}$ cm leads to%
\begin{equation*}
\ell ^{2}\left\vert \overline{p}\right\vert ^{2}\frac{\pi }{2}=8.88\times
10^{-20}
\end{equation*}%
much too small to be observable.

Next one computes the modifications to the quantum Coulomb spectrum arising
from (\ref{Ke3}). Because%
\begin{equation*}
\frac{1}{2}\left( \widehat{r}\cdot \overline{p}+\overline{p}\cdot \widehat{r}%
\right) =p_{r}=\frac{\hslash }{i}\frac{1}{r}\frac{\partial }{\partial r}r
\end{equation*}%
the radial equation becomes (in leading $\ell ^{2}$ order)%
\begin{equation*}
\left\{ \left( 1-\frac{2\epsilon \ell ^{2}me^{2}}{r}Z\right) \frac{d^{2}}{%
dr^{2}}-\frac{L\left( L+1\right) }{r^{2}}+\frac{2me^{2}}{\hslash ^{2}r}Z+%
\frac{2me}{\hslash ^{2}}E\right\} r\psi \left( r\right) =0
\end{equation*}%
One now considers the eigenstates of the unperturbed equation and treats the
term $\Delta =\frac{2\epsilon \ell ^{2}me^{2}}{r}Z\frac{1}{r}\frac{d^{2}}{%
dr^{2}}r$ acting on $\psi \left( r\right) $ as a perturbation. Because of
the $\frac{1}{r}$ factor in $\Delta $ one expects the largest effects to
occur for $s$ states. One obtains for the first and second $s$ states%
\begin{equation*}
\left\langle \psi _{1s},\Delta \psi _{1s}\right\rangle =-3\epsilon \ell ^{2}%
\frac{Z^{4}m^{3}e^{8}}{\hslash ^{4}}
\end{equation*}%
\begin{equation*}
\left\langle \psi _{2s},\Delta \psi _{2s}\right\rangle =-\frac{7}{16}%
\epsilon \ell ^{2}\frac{Z^{4}m^{3}e^{8}}{\hslash ^{4}}
\end{equation*}%
and denoting by $H_{0}$ the unperturbed Hamiltonian, with $\left\langle \psi
_{ns},H_{0}\psi _{ns}\right\rangle =-\frac{Z^{2}me^{4}}{2\hslash ^{2}n^{2}}$%
\begin{equation*}
\frac{\left\langle \psi _{1s},\Delta \psi _{1s}\right\rangle }{\left\langle
\psi _{1s},H_{0}\psi _{1s}\right\rangle }=6\epsilon \ell ^{2}\frac{%
Z^{2}m^{2}e^{4}}{\hslash ^{2}}
\end{equation*}%
\begin{equation*}
\frac{\left\langle \psi _{2s},\Delta \psi _{2s}\right\rangle }{\left\langle
\psi _{2s},H_{0}\psi _{2s}\right\rangle }=\frac{7}{2}\epsilon \ell ^{2}\frac{%
Z^{2}m^{2}e^{4}}{\hslash ^{2}}
\end{equation*}%
Of more significance is perhaps the mixing matrix element%
\begin{equation*}
\left\langle \psi _{2s},\Delta \psi _{1s}\right\rangle =-\frac{44}{27}%
\epsilon \ell ^{2}\frac{Z^{4}m^{3}e^{8}}{\hslash ^{4}}
\end{equation*}%
If $m=m_{e}$, the electron mass, and $\ell =10^{-19}$ cm%
\begin{equation*}
\frac{m_{e}^{2}\ell ^{2}e^{4}}{\hslash ^{2}}=0.357\times 10^{-21}
\end{equation*}%
However for muon atoms and large $Z$, this value is multiplied by a factor $%
\approx \left( 200\times Z\right) ^{2}$.

For other noncommutative corrections to the Coulomb problem refer to \cite%
{VilelaEPJ1} where, in particular, angular momentum effects were taken into
consideration.

\subsection{Phase-space volume effects}

The phase space contraction for $\epsilon =+1$ and the phase space expansion
for $\epsilon =-1$ have already been described in \cite{VilelaEPJ1} and \cite%
{VilelaEPJ2}. Here I simply rederive this result in the context of the
general representation (\ref{2.6}) and update the experimental perspectives.

Consider a particular space coordinate $x^{i}\circeq x$, $\overline{p}%
^{i}\circeq p$. Then%
\begin{equation}
x=i\left( \epsilon \ell ^{2}p\frac{\partial }{\partial \overline{\Im }}+%
\overline{\Im }\frac{\partial }{\partial p}\right)  \label{PS1}
\end{equation}%
The eigenstates of this operator are%
\begin{equation}
\left\vert x\right\rangle =\exp \left( -i\frac{x}{\ell }\tanh ^{-1}\left( 
\frac{\ell p}{\overline{\Im }}\right) \right)  \label{PS2}
\end{equation}%
for $\epsilon =+1$ and%
\begin{equation}
\left\vert x\right\rangle =\exp \left( -i\frac{x}{\ell }\tan ^{-1}\left( 
\frac{\ell p}{\overline{\Im }}\right) \right)  \label{PS3}
\end{equation}%
for $\epsilon =-1$.

To obtain the wave function of a momentum wave function on the $\left\vert
x\right\rangle $ basis $\left( \epsilon =+1\right) $ one projects by
integration on the $p,\overline{\Im }$ variables%
\begin{equation}
\left\langle x\mid k\right\rangle =\int J\left( p,\overline{\Im }\right) dpd%
\overline{\Im }e^{i\frac{x}{\ell }\tanh ^{-1}\left( \frac{\ell p}{\overline{%
\Im }}\right) }\delta \left( p-k\right)  \label{PS4}
\end{equation}%
$J\left( p,\overline{\Im }\right) $ being an integration density. To proceed
it is convenient to change variables to%
\begin{eqnarray}
x &=&i\ell \frac{\partial }{\partial \mu }  \notag \\
p &=&\frac{R}{\ell }\sinh \mu  \notag \\
\overline{\Im } &=&R\cosh \mu  \label{PS5}
\end{eqnarray}%
and convert (\ref{PS4}) into%
\begin{equation}
\left\langle x\mid k\right\rangle =\int dRd\mu e^{i\frac{x}{\ell }\mu
}\delta \left( \mu -\sinh ^{-1}\left( \frac{\ell k}{R}\right) \right) \delta
\left( R-1\right) =e^{i\frac{x}{\ell }\sinh ^{-1}\left( \ell k\right) }
\label{PS6}
\end{equation}%
The choice $R=1$ corresponds in (\ref{PS5}) to the choice of a particular
representation of the pseudo Euclidean algebra in two dimensions. It
corresponds to the choice of a density $J\left( p,\overline{\Im }\right) $%
\begin{equation*}
J\left( p,\overline{\Im }\right) =\delta \left( \overline{\Im }-\sqrt{1+\ell
^{2}p^{2}}\right)
\end{equation*}

For $\epsilon =-1$ a similar calculation leads to%
\begin{equation}
\left\langle x\mid k\right\rangle =e^{i\frac{x}{\ell }\sin ^{-1}\left( \ell
k\right) }  \label{PS7}
\end{equation}

The density of states is obtained from%
\begin{eqnarray*}
\frac{x+L}{\ell }\sinh ^{-1}\left( \ell k_{n}\right) &=&\frac{x}{\ell }\sinh
^{-1}\left( \ell k\right) +2\pi n \\
\frac{x+L}{\ell }\sin ^{-1}\left( \ell k_{n}\right) &=&\frac{x}{\ell }\sin
^{-1}\left( \ell k\right) +2\pi n
\end{eqnarray*}%
leading to%
\begin{equation}
\begin{array}{ccccc}
dn=\frac{L}{2\pi }\frac{dk}{\sqrt{1+\ell ^{2}p^{2}}} &  & \text{for} &  & 
\epsilon =+1 \\ 
dn=\frac{L}{2\pi }\frac{dk}{\sqrt{1-\ell ^{2}p^{2}}} &  & \text{for} &  & 
\epsilon =-1%
\end{array}
\label{PS8}
\end{equation}

For 3 dimensions \cite{VilelaEPJ2}%
\begin{equation}
\begin{array}{ccccc}
dn=\frac{V}{2\pi ^{2}}\frac{1}{\ell ^{2}}\frac{\left( \sinh ^{-1}\left( \ell
\left\vert p\right\vert \right) \right) ^{2}dk}{\sqrt{1+\ell ^{2}\left\vert
p\right\vert ^{2}}} &  & \text{for} &  & \epsilon =+1 \\ 
dn=\frac{V}{2\pi ^{2}}\frac{1}{\ell ^{2}}\frac{\left( \sin ^{-1}\left( \ell
\left\vert p\right\vert \right) \right) ^{2}dk}{\sqrt{1-\ell ^{2}\left\vert
p\right\vert ^{2}}} &  & \text{for} &  & \epsilon =-1%
\end{array}
\label{PS9}
\end{equation}

As discussed in \cite{VilelaEPJ1} \cite{VilelaEPJ2} the contraction or
expansion of the phase space has an impact on the cross sections of
elementary processes. For example for the cross section of the reaction%
\begin{equation*}
\gamma +p\rightarrow \pi +N
\end{equation*}%
of high energy proton cosmic rays, the contraction of phase space in the $%
\epsilon =+1$ case would allow cosmic ray protons of higher energies and
from further distances to reach the earth. From the calculations performed
in \cite{VilelaEPJ1}, one knows that the phase space suppression factor for
the photon pion production is a function of $\alpha =\omega _{\gamma
}^{\prime 2}\ell ^{2}$ , $\omega _{\gamma }^{\prime }$ being the photon
energy in the proton rest frame, the suppression being only appreciable if $%
\alpha \succeq 1$. In this case $\omega _{\gamma }^{\prime }=1.49\times
10^{-12}p_{P}^{0}$,$\ p_{P}^{0}$ being the proton energy. Therefore for this
reaction the effect would be very small for $\ell =O\left( 10^{-19}\text{cm}%
\right) $. In any case even for larger values of $\ell $ the GZK cutoff
would not be much changed, the main difference being a bigger size for the
GZK sphere, meaning that more cosmic ray protons from further distances
would be able to reach the earth.

A better place to look for the effects of this phase space suppression might
be an increase ($\epsilon =+1$) or decrease ($\epsilon =-1$) in particle
multiplicity in high energy reactions \cite{VilelaEPJ2}. This effect would
be important when $\ell k\sim O\left( 1\right) $, $k$ being the typical
reaction momentum. For $\ell =10^{-19}$cm this would occur for $k\approx
100-200$ TeV (in the range of the future FCC).

\subsection{Diffraction, interference and uncertainty relations}

Massless or massive wave equations in the noncommutative context are
solutions of \cite{VilelaJMP} 
\begin{equation}
\left[ p^{\mu },\left[ p_{\mu },\psi \right] \right] =0  \label{D1}
\end{equation}%
or%
\begin{equation}
\left[ p^{\mu },\left[ p_{\mu },\psi \right] \right] -m^{2}\psi =0
\label{D1a}
\end{equation}%
where $\psi $ may either be a scalar or a tensor element of the enveloping
algebra $U\left( \Re _{\ell ,\infty }\right) $ of the algebra $\Re _{\ell
,\infty }=\left\{ x^{\mu },M^{\mu \nu },\overline{p}^{\mu },\overline{\Im }%
\right\} $. They have a general solution%
\begin{equation}
\psi _{k}\left( x\right) =\exp \left( ik\cdot \frac{1}{2}\left\{ x,\overline{%
\Im }^{-1}\right\} _{+}\right)   \label{D2}
\end{equation}%
from which quantum fields may be constructed \cite{VilelaJMP} with $k^{2}=0$
or $m^{2}$. Notice that in (\ref{D2}) the $x^{\mu }$'s are simply algebra
elements, not the coordinates of the wave. Physical results are obtained
from the application of a \textit{state} to the algebra.

From the commutator $[\overline{p}^{\mu },x^{\nu }]=i\eta ^{\mu \nu }%
\overline{\Im }$ it also follows that the wave equations also have
factorized solutions%
\begin{equation}
\psi _{k}\left( x\right) =\prod_{\mu =0}^{3}\psi _{k^{\mu }}\left( x^{\mu
}\right)   \label{D3}
\end{equation}%
with%
\begin{equation}
\psi _{k^{\mu }}\left( x^{\mu }\right) =e^{i\frac{1}{2}k^{\mu }\left\{
x_{\mu },\overline{\Im }^{-1}\right\} _{+}}\hspace{1cm}\text{(fixed }\mu 
\text{)}  \label{D3a}
\end{equation}%
The factorized solutions may be used to study the diffraction problem. A
geometry is chosen with one or two long slits along the $x^{2}$ coordinate
and an incident wave along the third coordinate $\overrightarrow{k}=k%
\overrightarrow{e}_{3}$. The wave in the slit(s) will be a superposition of
localized states on the first space coordinate $x^{1}$, namely (for a single
slit) of width $2\Delta $ (in the $\overline{p}^{1},\overline{\Im }$
representation)%
\begin{equation}
\left\vert \chi _{1}\right\rangle _{+}=\int_{-\Delta }^{\Delta }dx^{1}e^{-i%
\frac{x^{1}}{\ell }\tanh ^{-1}\left( \frac{\ell p^{1}}{\overline{\Im }}%
\right) }  \label{D4}
\end{equation}%
for $\epsilon =+1$ and 
\begin{equation}
\left\vert \chi _{1}\right\rangle _{-}=\int_{-\Delta }^{\Delta }dx^{1}e^{-i%
\frac{x^{1}}{\ell }\tan ^{-1}\left( \frac{\ell p^{1}}{\overline{\Im }}%
\right) }  \label{D4a}
\end{equation}%
for $\epsilon =-1$. Therefore after passing the slit the wave is%
\begin{equation}
\left\vert \Psi \right\rangle _{k}=\psi _{k^{0}}\left( x^{0}\right) \int
d\xi \left\langle \psi _{\xi }\left( x^{1}\right) \right\vert \left. \chi
\right\rangle \psi _{\xi }\left( x^{1}\right) \psi _{\sqrt{k^{2}-\xi ^{2}}%
}\left( x^{3}\right)   \label{D5}
\end{equation}%
The projections $\left\langle \psi _{\xi }\left( x^{1}\right) \right\vert
\left. \chi \right\rangle $ of the slit state on the wave equation solution $%
\psi _{\xi }\left( x^{1}\right) $ will be computed in order $\ell ^{2}$. In
addition, because of the factorized nature of the solutions, one may use for
the operators $x^{1},p^{1}$ and $\overline{\Im }$, a subalgebra
representation instead of the general representation (\ref{2.6}), namely%
\begin{equation}
\begin{array}{llll}
\begin{array}{c}
x^{1}=x \\ 
p^{1}=\frac{1}{\ell }\sinh \left( \frac{\ell }{i}\frac{d}{dx}\right)  \\ 
\overline{\Im }=\cosh \left( \frac{\ell }{i}\frac{d}{dx}\right) 
\end{array}
&  &  & \epsilon =+1%
\end{array}
\label{D6}
\end{equation}%
\begin{equation}
\begin{array}{llll}
\begin{array}{c}
x^{1}=x \\ 
p^{1}=\frac{1}{\ell }\sin \left( \frac{\ell }{i}\frac{d}{dx}\right)  \\ 
\overline{\Im }=\cos \left( \frac{\ell }{i}\frac{d}{dx}\right) 
\end{array}
&  &  & \epsilon =-1%
\end{array}
\label{D6a}
\end{equation}%
Then in $O\left( \ell ^{2}\right) $ 
\begin{equation}
\overline{\Im }^{-1}=1-\epsilon \frac{1}{2}\left( \frac{\ell }{i}\right) ^{2}%
\frac{d^{2}}{dx^{2}}+\frac{5}{4}\left( \frac{\ell }{i}\right) ^{4}\frac{d^{4}%
}{dx^{4}}-\cdots   \label{D7}
\end{equation}%
and in the representation (\ref{D6})-(\ref{D6a}) the generalized localized
states are simply $\left\vert \chi _{1}\right\rangle =\delta \left(
x^{1}-\eta \right) $. The projection $\left\langle \psi _{\xi }\left(
x^{1}\right) \right\vert \left. \chi \right\rangle $ becomes 
\begin{eqnarray*}
\left\langle \psi _{\xi }\left( x^{1}\right) \right\vert \left. \chi
\right\rangle  &=&\frac{1}{2\pi }\int_{-\Delta }^{\Delta }e^{i\frac{1}{2}\xi
\left\{ x^{1},\overline{\Im }^{-1}\right\} }\delta \left( x^{1}-\eta \right)
d\eta  \\
&=&\frac{1}{\pi }e^{-\epsilon \frac{1}{4}\xi ^{2}\ell ^{2}}\frac{\sin \left(
\Delta \xi \left( 1-\epsilon \frac{1}{6}\xi ^{2}\ell ^{2}\right) \right) }{%
\xi -\epsilon \frac{1}{6}\xi ^{3}\ell ^{2}}+O\left( \ell ^{4}\right) 
\end{eqnarray*}%
Therefore the intensity of the diffracted wave at angle $\theta =\sin ^{-1}%
\frac{\xi }{k}$ is proportional to%
\begin{equation*}
\frac{\sin ^{2}\left( \Delta \xi \left( 1-\epsilon \frac{1}{6}\xi ^{2}\ell
^{2}\right) \right) }{\left( \Delta \xi \left( 1-\epsilon \frac{1}{6}\xi
^{2}\ell ^{2}\right) \right) ^{2}}
\end{equation*}

For two slits of width $2\Delta $ at a distance $2\Sigma $ the slit states
would be 
\begin{equation*}
\int_{-\Sigma -\Delta }^{-\Sigma +\Delta }+\int_{\Sigma -\Delta }^{\Sigma
+\Delta }dx^{1}\left\vert \chi _{1}\right\rangle
\end{equation*}%
with normalized diffracted intensity%
\begin{equation*}
\frac{\sin ^{2}\left( \Delta \xi \left( 1-\epsilon \frac{1}{6}\xi ^{2}\ell
^{2}\right) \right) \sin ^{2}\left( \Sigma \xi \left( 1-\epsilon \frac{1}{6}%
\xi ^{2}\ell ^{2}\right) \right) }{\Delta ^{2}\Sigma ^{2}\left( \xi \left(
1-\epsilon \frac{1}{6}\xi ^{2}\ell ^{2}\right) \right) ^{4}}
\end{equation*}

One sees that the effects of noncommutativity become important for $\xi \ell
\sim O\left( 1\right) $. For $\ell \sim 10^{-19}$cm this would be $\xi
\succsim 100$ TeV.

On the other hand, the noncommutativity of the spacetime coordinates implies
uncertainty relations on the simultaneous measurement of two space
coordinates or one space and one time coordinate. From%
\begin{equation*}
\lbrack x^{\mu },x^{\nu }]=-i\epsilon \ell ^{2}M^{\mu \nu }
\end{equation*}%
one obtains for $\Delta x^{\mu }=\left( \left\langle \psi \left\vert \left(
x^{\mu }\right) ^{2}\right\vert \psi \right\rangle -\left\langle \psi
\left\vert x^{\mu }\right\vert \psi \right\rangle ^{2}\right) ^{\frac{1}{2}}$%
\begin{equation*}
\Delta x^{\mu }\Delta x^{\nu }\geq \frac{1}{2}\ell ^{2}\left\langle \psi
\left\vert M^{\mu \nu }\right\vert \psi \right\rangle
\end{equation*}%
In particular one notices that there is no space-space uncertainty if $%
\left\vert \psi \right\rangle $ is spinless, but time-space uncertainty
leads to observable effects.

\section{Remarks and conclusions}

1. Approaching the question of noncommutative spacetime from the point of
view of deformation theory and the principle of stability of physical
theories, the first important observation is the independence of the length
scales of noncommutativity of the coordinates ($\ell $) and of the momenta ($%
\phi ^{-1}$). The scale of $\phi ^{-1}$ being associated to the
noncommutativity of translations is naturally associated to gravity and the
Planck length. However the scale of $\ell $ might be larger and it makes
sense to launch an experimental effort to find upper bounds or even the
value of this length scale. At the present time, in addition to a precise
analysis of phenomena of cosmological origin and a refinement of the
neutrino speed measurements, another possibility lies in phase space
modification effects on high energy colliders.

2. The estimates, performed here based on GRB data, point to values of $\ell 
$ in the range $10^{-19}-10^{-21}$ cm (or $\tau \in \left( 0.3\times
10^{-29}-0.3\times 10^{31}\text{ s}\right) $) favouring the higher part of
this range. However these estimates can only be taken as indicative or as
establishing upper bounds because of the large uncertainties on the
calculation of the spectral lags, on the statistics of the GRB pulses and on
the nature of the intrinsic spectral lags.  

2. The deformed $\Re _{\ell ,\infty }=\left\{ x^{\mu },M^{\mu \nu },%
\overline{p}^{\mu },\overline{\Im }\right\} $ algebra has also some
consequences concerning the structure of the fundamental interactions, in
particular those that are associated to connection-valued fields. In
particular the additional dimension in the differential algebra may imply
the existence of new interactions and states as well as a new extended
structure for the Dirac equation. These questions, not dealt with here,
because they have a less direct experimental verification, are described
elsewhere \cite{VilelaJMP} \cite{VilelaEPJ2} \cite{VilelaJMP}.

3. In the context of deformation theory, the transition from classical to
quantum mechanics appears as the stabilization of the unstable Poisson
algebra to the stable Moyal algebra. At the level of general nonlinear
functions of position and momentum the corresponding Hilbert space algebra
of operators is also stable, but the Heisenberg algebra itself%
\begin{equation*}
\left[ p,x\right] =-i\boldsymbol{1\hspace{2cm}}\hbar =1
\end{equation*}%
is not, because the c-number $\boldsymbol{1}$ commutes with both $p$ and $x$%
. Stabilization would then suggest a deformation to%
\begin{equation*}
\lbrack x,\boldsymbol{1}]=i\epsilon \ell ^{2}p;\hspace{2cm}[p,\boldsymbol{1}%
]=-i\epsilon ^{\prime }\phi ^{2}x
\end{equation*}%
and generalizing to $x^{\mu }$ and $p^{\mu }$ together with compatibility
with the Lorentz group would lead to the tangent space deformed algebra $\Re
_{\ell ,\infty }=\left\{ x^{\mu },M^{\mu \nu },\overline{p}^{\mu },\overline{%
\Im }\right\} $. In conclusion, in the framework of stable theories this
algebra is already implicit in the transition to quantum mechanics.

4. All calculations in the previous sections were carried out for the
algebra of the (noncommutative) tangent space limit $\phi ^{-1}\rightarrow
\infty $. When the full deformed algebra in (\ref{A1}) is used, the
noncommutativity of momenta in%
\begin{equation*}
\lbrack p^{\mu },p^{\nu }]=-i\epsilon ^{\prime }\phi ^{2}M^{\mu \nu }
\end{equation*}%
corresponds to the noncommutativity of spacetime translations. A similar
noncommutativity is what occur in a gravitational field. In this sense,
gravitation might also be considered an emergent property arising from
deformation theory and the principle of stability of physical theories.
Considering $\phi $ rather than the metric as defining gravitational field,
gravitation would be formulated as a $SO\left( 3,3\right) $ gauge theory 
\cite{VilelaIJTP}. An interesting consequence is that the gravitational
field might be a function of the Casimir invariants of $SO\left( 3,3\right) $
and not only of the energy-momentum tensor.

\section{Appendix: The general deformations of the Poincar\'{e}-Heisenberg
algebra}

The Poincar\'{e}-Heisenberg algebra%
\begin{equation}
\begin{array}{lll}
\lbrack M^{\mu \nu },M^{\rho \sigma }] & = & i(M^{\mu \sigma }\eta ^{\nu
\rho }+M^{\nu \rho }\eta ^{\mu \sigma }-M^{\nu \sigma }\eta ^{\mu \rho
}-M^{\mu \rho }\eta ^{\nu \sigma }) \\ 
\lbrack M^{\mu \nu },p^{\lambda }] & = & i(p^{\mu }\eta ^{\nu \lambda
}-p^{\nu }\eta ^{\mu \lambda }) \\ 
\lbrack M^{\mu \nu },x^{\lambda }] & = & i(x^{\mu }\eta ^{\nu \lambda
}-x^{\nu }\eta ^{\mu \lambda }) \\ 
\lbrack p^{\mu },x^{\nu }] & = & i\eta ^{\mu \nu } \\ 
\lbrack x^{\mu },x^{\nu }] & = & 0 \\ 
\lbrack p^{\mu },p^{\nu }] & = & 0%
\end{array}
\label{Ap1}
\end{equation}%
is not stable (rigid). Its 2-cohomology group has three nontrivial
generators, which lead to the following modified commutators\footnote{%
In the notation of Ref.\cite{Chrysso}, $\beta _{i}=q\alpha _{i}$ and $\Im
=qM $} \cite{Chrysso}%
\begin{equation*}
\begin{array}{lll}
\lbrack p^{\mu },x^{\nu }] & = & i\eta ^{\mu \nu }\Im +i\beta _{3}M^{\mu \nu
} \\ 
\lbrack x^{\mu },x^{\nu }] & = & i\beta _{2}M^{\mu \nu } \\ 
\lbrack p^{\mu },p^{\nu }] & = & i\beta _{1}M^{\mu \nu } \\ 
\lbrack x^{\mu },\Im ] & = & -i\beta _{2}p^{\mu }+i\beta _{3}x^{\mu } \\ 
\lbrack p^{\mu },\Im ] & = & i\beta _{1}x^{\mu }-i\beta _{3}p^{\mu } \\ 
\lbrack M^{\mu \nu },\Im ] & = & 0%
\end{array}%
\end{equation*}%
There is an instability cone at $\beta _{3}^{2}=\beta _{1}\beta _{2}$, but
for generic $\beta _{1},\beta _{2,}\beta _{3}$ all these algebras are rigid
and are isomorphic to either $SO\left( 1,5\right) $ or $SO\left( 2,4\right) $
or $SO\left( 3,3\right) $ (depending on the signs of $\beta _{1}$ and $\beta
_{2}$). For all these classes there is a representative with $\beta _{3}=0$,
which is exactly the deformation (\ref{A1}) obtained in \cite{VilelaJPA}.
The $\beta _{3}=0$ situation may always be obtained by a linear change of
coordinates in the algebra. The converse situation $\beta _{1}=\beta _{2}=0$
and $\beta _{3}\neq 0$ also mentioned in \cite{VilelaJPA} leads to $[x^{\mu
},x^{\nu }]=[p^{\mu },p^{\nu }]=0$ which does not seem to be physically
relevant, because at least the second commutator is expected to be different
from zero in the presence of gravity.

Here and elsewhere, I will be interpreting $x^{\mu }$ and $p^{\nu }$ as the
physical coordinates and momenta. In this sense I do not agree with the
criticism in \cite{Chrysso} about this choice, because not all observables
have to be extensive, only those that correspond to symmetry
transformations, in this case $M_{\mu \nu }$ and $p_{\mu }$.


\begin{thebibliography}{99}
\bibitem{Snyder} H. S. Snyder; \textit{Quantized spacetime}, Phys. Rev. 71
(1947) 38-41, \textit{The electromagnetic field in quantized spacetime},
Phys. Rev. 72 (1947) 68-71.

\bibitem{Yang} C. N. Yang; \textit{On quantized spacetime}, Phys Rev. 72
(1947) 874.

\bibitem{Douglas} M. R. Douglas and N. A. Nekrasov; \textit{Noncommutative
field theory}, Rev. Mod. Phys. 73 (2001) 977-1029.

\bibitem{Gaume} L. Alvarez-Gaum\'{e} and M. A. Vazquez-Mozo; \textit{General
properties of noncommutative field theory}, Nuclear Phys. B 668 (2003)
293-321.

\bibitem{Hinchliffe} I. Hinchliffe, N. Kersting and Y. L. Ma; \textit{Review
of the phenomenology of noncommutative geometry}, Int. J. Mod. Phys. A 19
(2004) 179-204.

\bibitem{Hossenfelder} S. Hossenfelder; \textit{Minimal length scale
scenarios for quantum gravity}, Living Rev. Relativity 16 (2013) 2-90.

\bibitem{AhluwaliaPLB} D. V. Ahluwalia; \textit{Quantum measurement,
gravitation, and locality}, Physics Letters B 339 (1994) 301-303.

\bibitem{Doplicher} S. Doplicher, K. Fredenhagen and J. E. Roberts;\ \textit{%
The quantum structure of spacetime at the Planck scale and quantum fields},
Commun. Math. Phys. 172 (1995) 187-220.

\bibitem{Amelino} G. Amelino-Camelia; \textit{Quantum-Spacetime Phenomenology%
}, Living Rev. Relativ. 16 (2013) 5.

\bibitem{VilelaSPTP} R. Vilela Mendes; \textit{The stability of physical
theories principle}, in "Beyond Peaceful Coexistence-The Emergence of Space,
Time and Quantum", I. Licata (ed.), pgs. 153-200, Imperial College Press
2016.

\bibitem{Andronov} A. Andronov and L. Pontryagin; \textit{Syst\`{e}mes
grossiers}, Dokl. Akad. Nauk. USSR 14 (1937) 247-251.

\bibitem{Smale} S. Smale; \textit{Differentiable dynamical systems}, Bull.
Am. Math. Soc. 73 (1967) 747-817.

\bibitem{Flato} M. Flato; \textit{Deformation view of physical theories},
Czech J. Phys. B32 (1982) 472-475.

\bibitem{Faddeev} L. D. Faddeev; \textit{On the Relationship between
Mathematics and Physics}, Asia-Pacific Physics News 3 (1988) 21-22 and in
\textquotedblright Frontiers in Physics, High Technology and
Mathematics\textquotedblright\ (ed. Cerdeira and Lundqvist) pp.238-246,
World Scientific, 1989.

\bibitem{VilelaJPA} R. Vilela Mendes; \textit{Deformations, stable theories
and fundamental constants}, J. Phys. A: Math. Gen. 27 (1994) 8091-8104.

\bibitem{VilelaPLA1} R. Vilela Mendes; \textit{Quantum mechanics and
noncommutative spacetime}, Phys. Lett. A 210 (1996) 232-240.

\bibitem{Chrysso} C. Chryssomalakos and E. Okon; \textit{Generalized quantum
relativistic kinematics: A stability point of view}, Int. J. Mod. Phys. D 13
(2004) 2003-2034.

\bibitem{Ahluwalia1} D. V. Ahluwalia-Khalilova; \textit{Minimal
spatio-temporal extent of events, neutrinos, and the cosmological constant
problem}, Int. J. Mod. Phys. D 14 (2005) 2151-2166.

\bibitem{Ahluwalia2} D. V. Ahluwalia-Khalilova; \textit{A freely falling
frame at the interface of gravitational and quantum realms}, Class. Quantum
Gravity 22 (2005) 1433-1450.

\bibitem{Morita} K. Morita; \textit{Discrete Symmetries in Lorentz-Invariant
Non-Commutative QED}, Progress Theor. Physics 110 (2003) 1003-1019.

\bibitem{VilelaJMP} R. Vilela Mendes; \textit{Geometry, stochastic calculus
and quantum theories in a noncommutative spacetime}, J. Math. Phys. 41
(2000) 156-186.

\bibitem{VilelaIJTP} R. Vilela Mendes; \textit{The geometry of
noncommutative spacetime}, Int. J. Theor. Phys. 56 (2017) 259--269.

\bibitem{VilelaEPJ2} R. Vilela Mendes; \textit{A laboratory scale
fundamental time?}, Eur. Phys. J. C 72 (2012) 2239.

\bibitem{VilelaMPLA} R. Vilela Mendes; \textit{An extended Dirac equation in
noncommutative spacetime}, Modern Physics Letters A 31 (2016) 1650089.

\bibitem{VilelaEPJ1} R. Vilela Mendes; \textit{Some consequences of a
noncommutative spacetime structure}, Eur. Phys. J. C 42 (2005) 445-452.

\bibitem{VilelaPLA3} R. Vilela Mendes; \textit{The deformation-stability
fundamental length and deviations from c}, Phys. Lett. A 376 (2012)
1823-1826.

\bibitem{OPERA} T. Adam et al.; \textit{Measurement of the neutrino velocity
with the OPERA detector in the CNGS beam}, JHEP 10 (2012) 093.

\bibitem{VilelaPLA2} E. Carlen and R. Vilela Mendes; \textit{noncommutative
spacetime and the uncertainty principle}, Phys. Lett. A 290 (2001) 109-114.

\bibitem{MINOS} P. Adamson et al.; \textit{Measurement of neutrino velocity
with the MINOS detectors and NuMI neutrino beam}, Phys. Rev. D 76 (2007)
072005.

\bibitem{SN1987A} K. Hirata et al.; \textit{Observation of a neutrino burst
from the supernova SN 1987a}, Phys. Rev. Lett. 58 (1987) 1490.

\bibitem{SN2} R. Bionta et al., \textit{Observation of a neutrino burst in
coincidence with supernova SN 1987a in the large magellanic cloud}, Phys.
Rev. Lett. 58 (1987) 1494.

\bibitem{SN3} M.J. Longo; \textit{Tests of relativity from SN 1987a}, Phys.
Rev. D 36 (1987) 3276.

\bibitem{AmelinoGRB} G. Amelino-Camelia , J. Ellis, N. E. Mavromatos, D. V.
Nanopoulos and S. Sarkar; \textit{Tests of quantum gravity from observations
of gamma-ray bursts}, Nature 393 (1998) 763-765.

\bibitem{Xu1} H. Xu and B. Q. Ma; \textit{Regularity of high energy photon
events from gamma ray bursts}, JCAP 1801 (2018) no.01, 050.

\bibitem{Ellis2} J. Ellis, R. Konoplich, N. E. Mavromatos, L. Nguyen, A. S.
Sakharov and E. K. Sarkisyan-Grinbaum; \textit{Robust constraint on Lorentz
violation using Fermi-LAT Gamma Ray Burst data}, arXiv:1807.00189.

\bibitem{Norris1} J. P. Norris, G. F. Marani and J. T. Bonnell; \textit{%
Connection between energy-dependent lags and peak luminosity in gamma-ray
bursts}, Astrophysical Journal 534 (2000) 248-257.

\bibitem{Norris2} J. P. Norris; \textit{Implications of the lag-luminosity
relationship for unified gamma-ray burst paradigms}, Astrophysical Journal
579 (2002) 386-403.

\bibitem{Ukwatta1} T. N. Ukwatta et al.; \textit{Spectral lags and the
lag-luminosity relation: an investigation with Swift Bat gamma-ray bursts},
Astrophysical Journal 711 (2010)\ 1073-1086.

\bibitem{Ukwatta2} T. N. Ukwatta et al.; \textit{The lag-luminosity relation
in the GRB source-frame: an investigation with Swift Bat bursts}, Mon. Not.
R. Astron. Soc. 419 (2012) 614--623.

\bibitem{Shao} L. Shao et al.; \textit{A new measurement of the spectral lag
of gamma-ray bursts and its implications for spectral evolution behaviors},
Astrophysical Journal 844 (2017) 126.

\bibitem{Zhang1} S. Zhang and B.-Q.\ Ma; \textit{Lorentz violation from
gamma-ray bursts}, Astroparticle Phys. 61 (2015) 108--112.

\bibitem{Bernardini} M. G. Bernardini et al.; \textit{Comparing the spectral
lag of short and long gamma-ray bursts and its relation with luminosity},
Mon. Not. R. Astron. Soc. 446 (2015) 1129-1138.

\bibitem{Bernardini2} M. G. Bernardini et al.; \textit{Limits on quantum
gravity effects from Swift short gamma-ray bursts}, Astronomy \&
Astrophysics 607 (2017) A121.

\bibitem{Wei} J.-J. Wei and X.-F. Wu; \textit{A further test of Lorentz
violation from the rest-frame spectral lags of gamma-ray bursts}, The
Astrophysical Journal 851 (2017) 127.

\bibitem{Blazar1} IceCube collaboration; \textit{Neutrino emission from the
direction of the blazar TXS 0506+056 prior to the IceCube-170922A alert,}
Science 361 (2018) 147-151.

\bibitem{Blazar2} IceCube collaboration; \textit{Multimessenger observations
of a flaring blazar coincident with high-energy neutrino IceCube-170922A, }%
Science 361 (2018) eaat1378.

\bibitem{Abbasi} R. Abbasi et al.; \textit{An absence of neutrinos
associated with cosmic-ray acceleration in }$\gamma $\textit{-ray bursts},
Nature 484 (2012) 351--354.

\bibitem{Adrian} S. Adri\'{a}n-Martinez et al. ; \textit{Search for muon
neutrinos from gamma-ray bursts with the ANTARES neutrino telescope using
2008 to 2011 data}, Astron. Astrophys. 559 (2013) A9.

\bibitem{Aartsen} M. G. Aartsen et al.; \textit{Search for Prompt Neutrino
Emission from Gamma-Ray Bursts with IceCube}, Astrophys. Journal 805 (2015)
L5.

\bibitem{Wei1} J.-J. Wei, B.-B. Zhang, L. Shao, X.-F. Wu and P. M\'{e}sz\'{a}%
ros; \textit{A new test of Lorentz invariance violation: The spectral lag
transition of GRB160625B}, Astrophys. J. Lett. 834 (2017) L13.

\bibitem{Wei2} J.-J. Wei, X.-F. Wu, B.-B. Zhang, L. Shao, P. M\'{e}sz\'{a}%
ros and V. A. Kostelecky; \textit{Constraining anisotropic Lorentz violation
via the spectral lag transition of GRB160625B}, Astrophys. J. 842 (2017) 115.

\bibitem{Ganguly} S. Ganguly and S. Desai; \textit{Statistical significacnce
of spectral lag transition in GRB160625B}, Astroparticle Phys. 94 (2017)
17-21.
\end{thebibliography}
\end{document}